\documentclass[11pt,a4paper]{article}
\usepackage{jcappub}
\usepackage{xspace}
\usepackage{color}

\usepackage{subcaption}
\PassOptionsToPackage{usenames,dvipsnames,svgnames}{xcolor}
\usepackage{tikz}
\usetikzlibrary{arrows,positioning,automata,shapes,snakes, matrix, fit}
\usepackage{color}
\usepackage{graphicx}
\usepackage{dcolumn}
\usepackage{bm}
\usepackage{hyperref}
\usepackage[normalem]{ulem}

\usepackage{booktabs}
\usepackage{amstext}
\usepackage{amssymb}
\usepackage{graphicx}
\usepackage{latexsym}
\usepackage{mathrsfs}
\usepackage{amsfonts}

\providecommand{\e}[1]{\ensuremath{\times 10^{#1}}}
\newcommand{\mpl}{M_\mathrm{Pl}}

\renewcommand{\arraystretch}{1.2}

\def\vev#1{ \left\langle #1 \right \rangle }
\def\|{\, | \,}
\def\P{\mathcal P}

\begin{document}

\begin{center}
\leftline{\small DESY-15-199}
\vskip -0.75cm
\end{center}

\title{Designing and testing inflationary models with Bayesian networks}

\author[1,2]{Layne C. Price,}
\author[3]{Hiranya V. Peiris,}
\author[4,5,6]{Jonathan Frazer,}
\author[2]{and Richard Easther}

\affiliation[1]{McWilliams Center for Cosmology, Department of Physics \\  Carnegie Mellon University, Pittsburgh, PA 15213, USA}
\affiliation[2]{Department of Physics,  University of Auckland \\ Private Bag 92019,  Auckland, New Zealand}
\affiliation[3]{Department of Physics and Astronomy,  University College London \\ London WC1E 6BT, UK}
\affiliation[4]{Deutsches Elektronen-Synchrotron DESY, Theory Group, 22603, Hamburg, Germany}
\affiliation[5]{Department of Theoretical Physics, University of the Basque Country, UPV/EHU \\ 48040 Bilbao, Spain}
\affiliation[6]{IKERBASQUE, Basque Foundation for Science, 48011 Bilbao, Spain}

\emailAdd{laynep@andrew.cmu.edu}
\emailAdd{h.peiris@ucl.ac.uk}
\emailAdd{jonathan.frazer@desy.de}
\emailAdd{r.easther@auckland.ac.nz}

\date{\today}

\abstract{
Even simple inflationary scenarios have many free parameters. Beyond the variables appearing in the inflationary action, these include dynamical initial conditions, the number of fields, and couplings to other sectors. These quantities are often ignored but cosmological observables can depend on the unknown parameters.  We use Bayesian networks to account for a large set of inflationary parameters, deriving \emph{generative models\/} for the primordial spectra that are conditioned on a hierarchical set of prior probabilities describing the initial conditions, reheating physics, and other free parameters. We use $N_f$--quadratic inflation as an illustrative example, finding that the number of $e$-folds $N_*$ between horizon exit for the pivot scale and the end of inflation is typically the most important parameter, even when the number of fields, their masses and initial conditions are unknown, along with possible conditional dependencies between these parameters.
}

\maketitle

\section{Introduction}

Precise measurements of the cosmic microwave background from \emph{Planck}~\cite{Ade:2013rta,Ade:2015lrj} and WMAP~\cite{Hinshaw:2012aka} have put tight constraints on  estimated values of the free parameters of simple inflationary models. However these analyses have only been performed for the simplest possible models. For instance,  additional scalar fields, nontrivial couplings to gauge fields or the gravitational sector, or varying the initial conditions all introduce new variables into the model. 

Parameters associated with these variables are  implicitly set to zero in these treatments, and including them explicitly would make extracting constraints on the models considerably  more challenging.  These parameters can alter the predicted primordial density perturbations, and realistic treatments of these scenarios must account for this uncertainty in the \emph{a priori} structure of the underlying model. This uncertainty should then be consistently propagated into the predictions for the early universe and enfolded into the comparisons with data. In this paper we address this challenge by introducing a Bayesian {\sl hierarchical modelling} scheme for inflationary physics, which incorporates theoretical and statistical uncertainties within a unified framework.

\begin{figure}
  \centering
  \begin{tikzpicture}[
    myscope/.style={node distance=1em and 0em},
    mymatrix/.style={matrix of nodes, nodes=block,
      column sep=2em,
      row sep=1cm},
    block/.style={draw=white, thick, fill=blue!10, rounded corners,
      minimum width=6em,
      minimum height=2em},
    vhilit/.style={draw=red, thick, dotted,
      inner sep=1em},
    hhilit/.style={draw=black, thick,
      inner xsep=2em,
      inner ysep=.5em},
    line/.style={thick, -latex, shorten >= 2pt}
    ]

    \matrix[mymatrix] (mx) {
      $\mathcal O(2+)$ Hyperparameters: $\xi$ \\
      $\mathcal O(1)$ Hyperparameters: $\xi$ \\
      Model parameters: $\theta$ \\
      Data/Observables: $D$, $x$ \\
    };

    \draw[line] (mx-1-1) -- (mx-2-1);
    \draw[line] (mx-2-1) -- (mx-3-1);
    \draw[line] (mx-3-1) -- (mx-4-1);

  \end{tikzpicture}
  ~~~~~~~
  ~~~~~~~
  \begin{tikzpicture}[>=stealth',shorten >=1pt,node distance=1.8cm and 1.5cm,on grid,
    initial/.style    ={},
    block/.style={draw=black, thick, rounded corners,
    minimum width=2.5em,
    minimum height=2.5em},
  ]

  \tikzset{every node/.style={fill=GreenYellow}}
    \node[block]          (0)                         {$\lambda$};
  \tikzset{every node/.style={fill=Dandelion}}
  \node[block]          (1) [below right = of 0]                         {$r$};
  \node[block]          (6) [below left = of 0]                         {$n_s$};
  \tikzset{every node/.style={fill=white}}
  \node[block]          (2) [above left = of 0]                         {$a$};
  \node[block]          (3) [above right = of 0]                         {$b$};
  \tikzset{every node/.style={fill=Tan}}
  \node[block]          (4) [above = of 2]                         {$a_0$, $a_1$};
  \node[block]          (5) [above = of 3]                         {$b_0$, $b_1$};

  \tikzset{mystyle/.style={->,double=CadetBlue,dashed}}
  \path (2)     edge [mystyle]     (0);
  \path (3)     edge [mystyle]     (0);
  \path (4)     edge [mystyle]     (2);
  \path (5)     edge [mystyle]     (3);

  \tikzset{mystyle/.style={->,double=orange}}
  \path (0)     edge [mystyle]     (1);
  \path (0)     edge [mystyle]     (6);

  \end{tikzpicture}

  \caption{\label{fig:diagram} (\emph{Left}) A schematic relationship between the parameters and hyperparameters in a Bayesian network.  The arrows indicate that the target node depends on the source node. (\emph{Right}) A Bayesian network for a single-field, slow-roll inflation model with inflaton self-couplings $\lambda$.  The nodes correspond to observables (orange), model parameters (green), hyperparameters (white), and higher order hyperparameters (brown).  The dashed, blue arrows show a dependency through conditional probabilities; the solid, orange arrows indicate a deterministic dependency.  For this toy model the quasi-observables $n_s$ and $r$ are uniquely determined given the self-couplings $\lambda$, which are drawn uniformly in the range $a<b$ and the limits of this range are drawn uniformly from  $a_0<a_1$ and $b_0<b_1$.}
\end{figure}
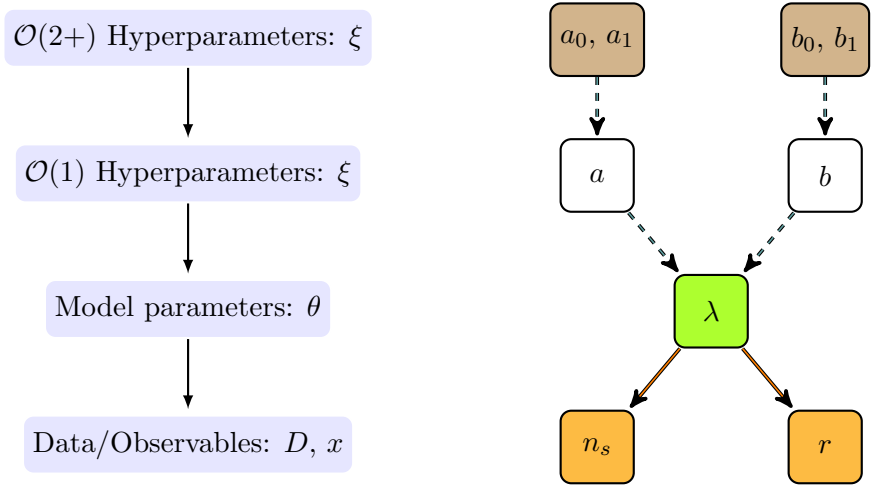

Hierarchical models have been used in other cosmological contexts, including inference problems with Type Ia supernovae data~\cite{Mandel:2009xr,March:2011xa,Mandel:2014aea,Rubin:2015rza,Shariff:2015yoa}, detection algorithms for spatially-localized features in the cosmic microwave background (CMB)~\cite{Feeney:2012hj}, and the analysis of cosmic shear data~\cite{Alsing:2015zca}. Our approach treats each inflationary parameter as a random variable with an associated prior probability distribution. These priors may subsequently depend on other model parameters or \emph{hyperparameters}, which can also be randomly distributed.  For example, the mean and variance are hyperparameters for normally-distributed model parameters.\footnote{In Sect.~\ref{sect:ex_nquad} the distribution of scalar masses $P(m^2 \| \beta)$ has a hyperparameter $\beta$ that is the ratio of the number of axions to the total number of moduli~\cite{Easther:2005zr}.} The data or predicted observables are generated stochastically by fixing the outermost hyperparameters in the hierarchy and sampling the chain of dependent prior probabilities.
These complicated dependency chains are treated as  \emph{Bayesian networks} in Sect.~\ref{sect:hier}, which give an easily visualized graphical depiction of the nested parameters.  A toy Bayesian network for a simple inflation model is shown in Fig.~\ref{fig:diagram}.

Our methodology can be summarized as follows:
\begin{enumerate}

  \item  \emph{Outline hierarchical dependencies.---}  Carefully describe the nested set of conditional dependencies between the parameters and hyperparameters in the inflation model and write down a graphical Bayesian network.  Subtly different choices can give substantially different predictions so this is a crucial step.
  
  \item  \emph{Define prior probabilities \& hyperprior probabilities.---}  Treat every parameter and hyperparameter as a random variable with an associated prior probability distribution; these distributions are related by the details of the Bayesian network.

  \item  \emph{Sample the Bayesian network.---}  Fix the value of the outermost hyperparameters.  Sample the prior probabilities for each subsequent layer of (hyper)parameters in an order set by their conditional dependencies.  Calculate the prediction using the final set of model parameters.  Repeat to build a large sample from which a probability distribution can be estimated.

\end{enumerate}

These hierarchical  methods are applicable to any inflation model, but are especially useful with scenarios with multiple active fields or couplings between the inflaton and other sectors.
We illustrate this approach with three variants of $N_f$--quadratic inflation, where the Bayesian networks exhibit varying degrees of complexity.
We present results for the  distribution of the tensor-to-scalar ratio $r$, the spectral tilt $n_s$, and scalar running $\alpha_s$ in Sect.~\ref{sect:ex_nquad}, with the standard definitions for the primordial spectra which are summarized in Sect.~\ref{ssect:nquad_basics}.

This model predicts a tensor-to-scalar ratio of $r \sim \mathcal O(10^{-1})$,\footnote{This estimate assumes that observables do not change during the post-inflationary epochs, although the evolution may not be adiabatic at the end of inflation for this model. Throughout this paper we ignore this potentially important dependence on post-inflationary physics, although in principle it could be included as an additional layer in the Bayesian network.}
which is in tension with the joint \emph{Planck} and BICEP2/\emph{Keck Array} results for the $\Lambda$CDM+$r$ model~\cite{Ade:2013zuv,Ade:2013rta,Mortonson:2014bja,Flauger:2014qra,Ade:2015tva,Ade:2015xua}, although it is more consistent with an extended $\Lambda$CDM scenario~\cite{DiValentino:2015ola}.
However, $N_f$--quadratic inflation is a useful example in that it is sufficiently complex to have a non-trivial Bayesian network, while still allowing an intuitive understanding of the inflationary dynamics.
Importantly, the techniques we use here apply to any inflation model, including those that provide a better fit to the CMB data, \emph{e.g.}, single-field models or models with non-canonical kinetic energies or non-minimal couplings to the gravity sector.

We confirm previous indications \cite{Easther:2013rva,Wenren:2014cga,Price:2014ufa} that $N_f$--quadratic models are largely insensitive to  the exact particle masses, initial conditions, and the total number of fields and conclude that the unknown reheating mechanism is the dominant source of uncertainty in these models, evidenced by a strong dependency on the  number of $e$-folds $N_*$ occurring between the moment at which the pivot scale $k_*$ leaves the horizon and the end of inflation.

This paper is organized as follows: in Section~\ref{sect:hier} we introduce generative hierarchical models and Bayesian networks.  In Section~\ref{sect:ex_nquad} we explore three Bayesian networks for $N_f$--quadratic inflation and give the slow-roll predictions for the spectral index $n_s$, the scalar running $\alpha_s$, and the tensor-to-scalar ratio $r$ at the end of inflation.  In Section~\ref{sect:concl} we summarize our results.

\section{Generative hierarchical models \& Bayesian networks}
\label{sect:hier}

To compare an inflation scenario's prediction to experimental data we need to know (A) how to make a prediction for the data with fixed model parameters and (B) the distribution from which these parameters are to be drawn.
We divide a model's parameters and their prior probabilities into three qualitative classes:
\begin{enumerate}

  \item \emph{Model parameters \& model priors---} Parameters $\theta$ for an inflation model that directly affect the prediction, and prior probabilities $P(\theta)$ that encode how we expect $\theta$ to be distributed based on theoretical or empirical knowledge.  For example, self couplings $\lambda$, masses $m^2$, the number of fields $N_f$, or the number of $e$-folds $N_*$ after the pivot scale leaves the horizon could all be model parameters:
    \begin{equation}
      \theta \sim \left\{\lambda, m^2, N_f, N_*, \dots \right\}.
      \label{eqn:XXX}
    \end{equation}
    The model parameters traditionally define the physics.

  \item \emph{Hyperparameters \& hyperpriors---}  Variables $\xi$ that define the model priors $P(\theta \| \xi)$.  For instance, if a model parameter $\theta$ is expected to be normally distributed as $\theta \sim \mathcal N(\mu, \sigma)$, then the mean $\mu$ and standard deviation $\sigma$ are hyperparameters:
    \begin{equation}
      \xi \sim \left\{ \mu, \sigma, \dots \right\}.
      \label{eqn:XXX}
    \end{equation}
    Hyperparameters are distributed according to hyperpriors $P(\xi)$.  The hyperparameters/hyperpriors statistically generate realizations of the inflationary scenarios.
A nested set of hyperparameters may be required if the lower order hyperparameters are inherently stochastic, if their true value is unknown, or if we wish to marginalize over prior probability distributions in order to assess the robustness of the Bayesian network.

  \item \emph{Data \& likelihoods---}  The predicted data $D$ are treated as random variables in the model with some likelihood $P(D \| \theta, \xi)$.
    In place of simulating experimental data we can replace $D$ with proxy quasi-observables $x$, such as parametrizations of the primordial spectra
    \begin{equation}
      x \sim \left\{ A_s, n_s, r, f_\mathrm{NL}, \dots \right\}.
      \label{eqn:XXX}
    \end{equation}
    Throughout, we refer to data $D$ or quasi-observables $x$ interchangeably.

\end{enumerate}

By sampling the prior and hyperprior distributions we  obtain stochastic predictions for the data and/or quasi-observables in the model, which are conditioned on the prior probabilities.  This defines a \emph{generative model} for the primordial parameters:\footnote{Although the generative model could also directly predict the CMB data, in this paper we restrict attention to proxy quasi-observables.}
\begin{equation}
  \left\{ m^2, N_*, \dots \right\} \quad \xrightarrow{\mathrm{generative}} \quad \left\{A_s, n_s, r, f_\mathrm{NL}, \dots \right\}.
  \label{eqn:XXX}
\end{equation}
We specify generative inflation models using Bayesian networks, which outline conditional dependencies between parameters, hyperparameters, priors, and the predicted data by a directed acyclic graph.
Figure~\ref{fig:diagram} shows a schematic depiction of a Bayesian network along with a toy network for an inflationary model where the spectral index and the tensor-to-scalar ratio are simple functions of the inflaton self-couplings $\lambda$.
The arrows indicate conditional dependence between the target and source node. The network is directed so that the probability distribution for an interior variable in the model is determined exclusively from the higher levels in the graph.
The interior parameters and the data are generated according to this chain of probabilistic dependencies.
Since the graph is  acyclic, there will be a set of outermost parameters $\xi_\mathrm{out}$, which we set to fixed values. 
Repeated sampling of the generating distribution results in a sample of predicted data that follow the marginalized \emph{a priori} distribution
\begin{equation}
  P(D \| \xi_\mathrm{out}) = \int P(D \| \theta, \xi, \xi_\mathrm{out}) P(\theta, \xi \| \xi_\mathrm{out}) \; d \theta  \;d \xi.
  \label{eqn:XXX}
\end{equation}
The prior $P(\theta, \xi \| \xi_\mathrm{out})$ can be factorized into conditional probabilities that define the structure of the Bayesian network.  Finally, varying the  $\xi_\mathrm{out}$ gives a qualitative measure of the sensitivity of the model's predictions to the values of the hyperparameters.

\section{Example: $N_f$--quadratic inflation}
\label{sect:ex_nquad}

\subsection{The model parameters and observables}
\label{ssect:nquad_basics}

\begin{figure}
\centering
\includegraphics[width=1.0\textwidth]{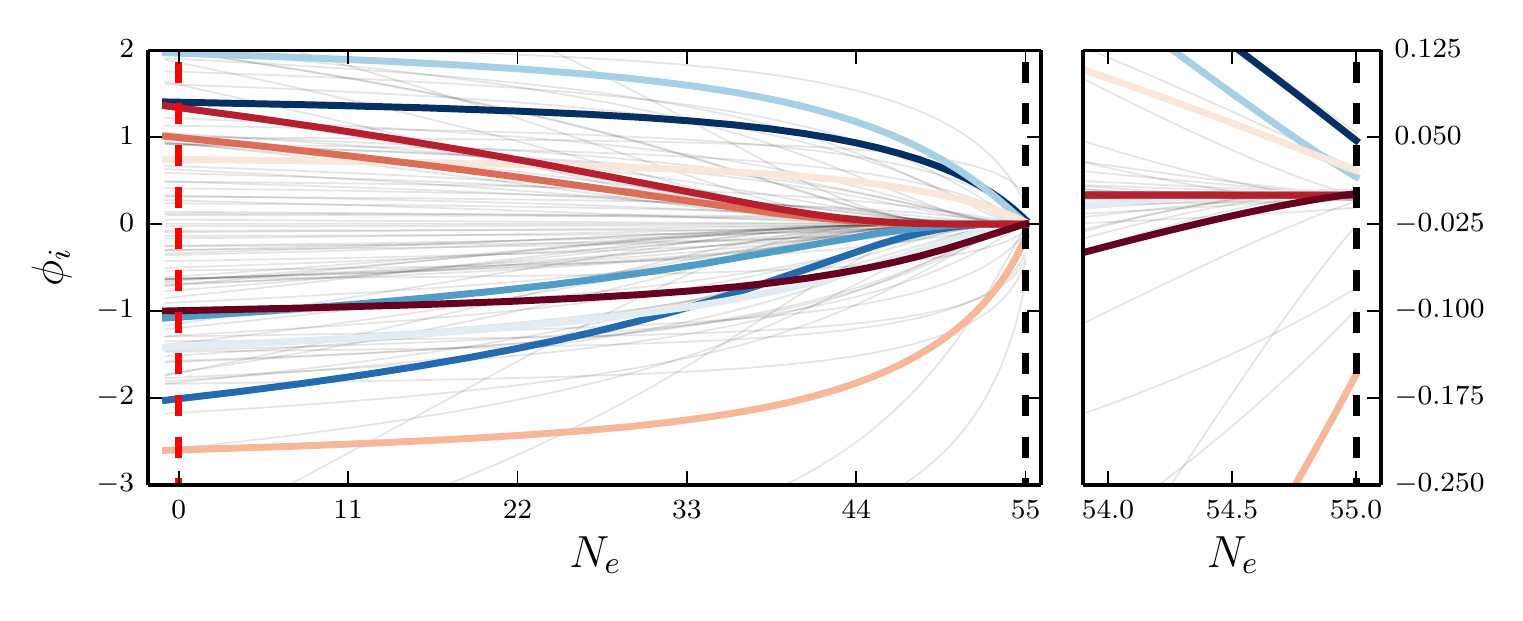}
\caption{Evolution of the fields $\phi_i$ as a function of $e$-folds $N_e$ after the pivot scale leaves the horizon at $N_e =0$, with the $N_f$--quadratic potential~\eqref{eqn:Nquad} and $N_* = 55$.  Masses are set by the Mar\v{c}enko-Pastur distribution~\eqref{eqn:MP} and 100 fields are plotted, with 10 randomly chosen light fields emphasized in color.  (\emph{Left})  Initial conditions are set near the horizon crossing surface (\emph{dashed red line}). (\emph{Right}) Zoom-in near the end of inflation surface (\emph{dashed black line}), demonstrating genuinely multifield behavior for the entirety of the inflationary period.}
\label{fig:field_evol}
\end{figure}

\begin{table}
  \centering
  \renewcommand{\arraystretch}{1.2}
  \setlength{\arraycolsep}{5pt}
  \begin{tabular}{ p{2.0cm} | p{8cm} | p{3cm} }
    Model & Description & Reference \\
    \hline
    BayesNet A               & A simple network with minimal dependencies & Fig.~\ref{fig:DAG_nquad1} \& Sect.~\ref{ssect:pheno} \\
    BayesNet B               & The prior for $N_f$ maximizes the Shannon entropy & Fig.~\ref{fig:DAG_nquad1} \& Sect.~\ref{ssect:maxent} \\
    BayesNet C               & Incorporates a dependence on the initial energy and requires sub-Planckian field displacements & Fig.~\ref{fig:DAG_nquad2} \& Sect.~\ref{ssect:hep} \\
  \end{tabular}
  \caption{A legend for the three example $N_f$--quadratic models.}
  \label{table:legend}
\end{table}

\paragraph{The potential.---}
In Sects.~\ref{ssect:pheno}--\ref{ssect:hep} we consider several specific scenarios that illustrate the utility of the hierarchical networks.  We summarize the models in Table~\ref{table:legend}.   This methodology does not impose any constraints on the nature of the inflationary Lagrangian and applies equally well to complicated and simple scenarios.  However, in most single-field models the relevance of various parameters tends to be obvious, so we focus on  more complex models.

Consider the $N_f$--quadratic inflation model, which has $N_f$ massive fields, a canonical kinetic energy term, and the sum-separable potential
\begin{equation}
  V(\phi_1,\dots,\phi_{N_f})\equiv \sum_i V_i(\phi_i) = \frac{1}{2} \sum_i^{N_f} m_i^2 \phi_i^2.
  \label{eqn:Nquad}
\end{equation}
If the number of fields $N_f$ is large this potential can realize assisted inflation~\cite{Liddle:1998jc,Dimopoulos:2005ac}, where each field undergoes a sub-Planckian field displacement while  generating observable gravitational wave backgrounds, with $r \gtrsim 0.1$.  If we assume  each field has an initial value $|\phi_i| <1$ and all $m_i^2 >0$ this potential is the lowest-order description for a set of canonical fields rolling into a minimum.
The  curvature spectra derived from the potential~\eqref{eqn:Nquad} have been well studied~\cite{Liddle:1998jc,Kanti:1999ie,Easther:2005zr,Dimopoulos:2005ac,Kim:2006ys,Kim:2006te,Kim:2007bc,Kim:2011jea} and marginalized predictions for the power spectrum were obtained in Refs.~\cite{Frazer:2013zoa,Easther:2013rva,Wenren:2014cga,Price:2014ufa}, but no complete generative models have  been developed and analyzed for this scenario.

Even if all of the model parameters are fixed, observables can still depend on the initial states $\phi_{i,0}$ of the homogeneous background fields, particularly in multifield models. For generic initial conditions and masses the potential~\eqref{eqn:Nquad} has genuinely multifield behavior throughout the entirety of inflation, as the field evolution does not generally reach a single-field attractor.  The trajectory of the field values as a function of the number of $e$-folds of expansion $N_e$ during inflation is shown in Fig.~\ref{fig:field_evol}, for a model with fields of different masses.  At the end of inflation there are typically two or more active fields.

By contrast, a single-field model can use the monotonically decreasing field value $\phi$ as a clock during inflation, and within the slow-roll approximation a given field value can be uniquely associated with the number of $e$-folds that will occur before the end of inflation.  Multifield models only have a monotonically decreasing energy density $\rho$ and the values of  individual fields $\phi_i$ depend on the initial values.
This initial state is unknown, so the \emph{a priori} probability distribution of initial field values  induces a probability distribution for the model's predictions.  The inherently stochastic predictions of multifield models are well described by the generative modelling methodology of Sect.~\ref{sect:hier}.

\paragraph{Methodology.---}
We will focus on the predictions of $N_f$--quadratic inflation  for the primordial power spectrum of curvature perturbations $\P_{\zeta} (k)$ at the end of inflation.  While higher order correlators could also be computed either analytically or numerically, the two-point function is most easily constrained by data.  Furthermore, the $N_f$--quadratic model generally predicts a low level of non-Gaussianity well within current  bounds from \emph{Planck}~\cite{Ade:2013ydc,Ade:2015ava}, \emph{e.g.}, $|f_\mathrm{NL}| \sim 1/N_* \sim \mathcal O(10^{-2})$.

We will evaluate the spectral index $n_s$ of the scalar spectrum, the running of the spectral index $\alpha_s$, and the tensor-to-scalar ratio $r$ using the $\delta N$ formalism, which relates curvature perturbations $\zeta$ on constant density hypersurfaces to field perturbations $\delta \phi^i$ on flat hypersurfaces as
\begin{equation}
  \zeta = \frac{\partial N_e}{\partial \phi_i} \delta \phi^i + \mathcal O(\delta \phi_j^2),
  \label{eqn:dN}
\end{equation}
where $N_e$ is the number of $e$-folds between the two hypersurfaces, given by
\begin{equation}
N_e = - \sum_i^{N_f} \int_{\phi_{i,0}}^{\phi_{i,c}} \left( \frac{V_i}{V_i'} \right) \, d\phi_i
  \label{eqn:Npiv}
\end{equation}
assuming the slow-roll conditions.
If the sub-horizon evolution of the field perturbations gives no cross-correlations $\vev{\delta \phi_i \delta \phi_j^*}$ at horizon exit, then
any sum-separable inflation model like Eq.~\eqref{eqn:Nquad} predicts a spectral index of~\cite{Vernizzi:2006ve,Battefeld:2006sz,Frazer:2013zoa}
\begin{equation}
  n_s -1 = - 2 \epsilon_* - 4
  \left[ 1- \sum_i^{N_f} \frac{\eta_i^* u_i^2}{2 \epsilon_i^*} \right]
  \left[ \sum_i^{N_f} \frac{u_i^2}{\epsilon_i^*} \right]^{-1},
  \label{eqn:ns_sr}
\end{equation}
a tensor-to-scalar ratio of
\begin{equation}
  r = 16 \left[ \sum_i^{N_f} \frac{u_i^2}{\epsilon_i^*} \right]^{-1},
  \label{eqn:r_sr}
\end{equation}
and a running of the spectral index of
\begin{align}
  \label{eqn:alpha_sr}
  \alpha_s
  =& -8 \epsilon_*^2
  + 4 \sum_i^{N_f} \epsilon_i^{*} \eta_i^*
  - 16 \left[ \sum_i^{N_f} \frac{u_i^2}{\epsilon_i^*} \right]^{-2} \left[ 1 - \sum_i^{N_f} \frac{\eta_i^* u_i^2}{2 \epsilon_i^*}  \right]^2 \\
  &-8 \left[ \sum_i^{N_f} \frac{u_i^2}{\epsilon_i^*} \right]^{-1} \sum_i^{N_f} \eta_i^* u_i \left[1 - \frac{\eta_i^* u_i^2}{2 \epsilon_i^*} \right] \notag \\
  &+ \frac{4}{\epsilon_*} \left[ \sum_i^{N_f} \frac{u_i^2}{\epsilon_i^*} \right]^{-1} \sum_i^{N_f} \frac{\eta_i^* u_i^2}{2 \epsilon_i^*}
  - 2 \left[ \sum_i^{N_f} \frac{u_i^2}{\epsilon_i^*} \right]^{-1} \sum_i^{N_f} \frac{\xi_i^* u_i^2}{\epsilon_i^*}.
  \notag
\end{align}
We define the slow-roll parameters as
\begin{equation}
  \epsilon \equiv \sum_i^{N_f} \epsilon_i \equiv \sum_i^{N_f} \frac{1}{2} \left( \frac{V_i'}{V} \right)^2,
  \qquad
  \eta_i \equiv \frac{V_i''}{V},
  \qquad \mathrm{and} \qquad
  \xi_i^2 \equiv \frac{V_i' V_i'''}{V^2},
  \label{eqn:XXX}
\end{equation}
where $V_i' \equiv dV_i/d\phi_i$ and similarly for higher derivatives.
We have also used the functions
\begin{equation}
  u_i \equiv \frac{V_i^* + Z_i^c}{V}
  \label{eqn:XXX}
\end{equation}
and
\begin{equation}
  Z_i^c \equiv \frac{1}{\epsilon_c} \sum_j^{N_f} V_j^c \left[ \epsilon_i^c - \left(\sum_k^{N_f} \epsilon_k^c \right) \delta_{ij} \right],
  \label{eqn:Z}
\end{equation}
where $\delta_{ij}$ is the Kronecker delta.
Quantities labeled by $(*)$  are evaluated as modes leave the horizon with $k=aH$ and quantities labelled by $(c)$ in Eq.~\eqref{eqn:Z} are evaluated at the end of inflation.
Ignoring possible cross-correlations at horizon crossing is a very good approximation for this model, as shown in comparison to exact numerical results in Refs.~\cite{Easther:2013rva,Price:2014ufa}.
In this paper we will describe $n_s$, $\alpha_s$, and $r$ only near the pivot scale $k_*$.

We evaluate Eqs.~\eqref{eqn:ns_sr}--~\eqref{eqn:alpha_sr} numerically with the publicly available package \textsc{MultiModeCode}~\cite{Mortonson:2010er,Easther:2011yq,Norena:2012rs,Price:2014xpa}, which numerically evolves the Klein-Gordon equations
\begin{equation}
  \frac{d^2 \phi_i}{d N_e^2} + \left(3 - \epsilon \right) \frac{d\phi_i}{dN_e} + \frac{1}{H^2} \frac{\partial V}{\partial \phi_i} = 0
  \label{eqn:kg}
\end{equation}
from some initial state $\{\phi_{i,0}, \phi_{i,0}'\}$ until the end of inflation when $\epsilon \equiv \sum_i \epsilon_i = 1$.  Given masses $m_i$ and the number of $e$-folds $N_*$ between when the pivot scale $k_*$ leaves the horizon and the end of inflation, the field values at horizon crossing $\phi_{i,*}$ and the functions $u_i$ and $Z_i^c$ can be calculated numerically.  The details of this procedure can be found in Ref.~\cite{Price:2014xpa}.

\paragraph{The horizon crossing approximation (HCA).---}
We can simplify the expressions for $u_i$ and $Z_i^c$ if we ignore any contributions to the observables from the end-of-inflation surface, which sets $Z_i^c \to 0$.  This is the \emph{horizon crossing approximation} (HCA) \cite{Vernizzi:2006ve,Kim:2006te}, which has been shown to be a reliable estimate for $n_s$ and $r$, even in the many-field limit $N_f \gg 1$ of $N_f$--quadratic inflation~\cite{Frazer:2013zoa,Price:2014ufa,Easther:2013rva}.  We discuss this further in Sect.~\ref{ssect:pheno}.
For chaotic potentials like Eq.~\eqref{eqn:Nquad}, the number of $e$-folds between horizon exit and the end of inflation from Eq.~\eqref{eqn:Npiv} is
\begin{equation}
  N_*  = \frac{1}{4} \sum_i^{N_f} \left[ \phi_{i,0}^2 - \phi_{i,c}^2 \right] \approx \frac{1}{4} \sum_i^{N_f} \phi_{i,0}^2,
  \label{eqn:Ntot}
\end{equation}
where we have ignored the end-of-inflation contribution.
Consequently, an arbitrarily large amount of inflation can be generated by setting the initial conditions $\phi_{i,0}$ to larger values.
This sets a close relationship between $N_e$ and its derivatives:
\begin{equation}
  \sum_i \left(\frac{\partial N_*}{\partial \phi_i} \right)^2 = N_*.
  \label{eqn:dN_hca}
\end{equation}
If the two-point correlation functions of the field perturbations at horizon crossing are diagonal $\vev{\delta \phi_i \delta \phi_j^*} \sim \delta_{ij} $, then the field-field power spectrum is related directly to the curvature power spectrum in Eq.~\eqref{eqn:dN} by $N_*$.

Substituting $V$ from Eq.~\eqref{eqn:Nquad} into Eq.~\eqref{eqn:r_sr}, we get
\begin{align}
  \label{eqn:Nquad_r}
&{\cal P_{\zeta}}= \frac{H^{2}}{4\pi^{2}}N_{*}, \quad && n_{s}-1= -2\epsilon-\frac{1}{N_{*}}, \\
&\alpha_s = -8\epsilon^2-\frac{1}{N_{*}^2}+4\epsilon_{i}\eta_{i}, \quad && r =\frac{8}{N_{*}} \, . \nonumber
\end{align}
Importantly, $n_s$ and $\alpha_s$ depend directly on the background field values at horizon crossing $\phi_i^*$, the masses of each field $m_i^2$, and the number of fields $N_f$ through the slow-roll parameters $\epsilon_i$ and $\eta_i$.  However, we will find that the relationship Eq.~\eqref{eqn:dN_hca} ensures that $N_*$ remains the most important parameter in this model for predicting $n_s$ and $\alpha_s$.

We will not use the HCA when computing observables for the models in Sects.~\ref{ssect:pheno}--\ref{ssect:hep}.  However, we  assume that the chain of dependencies revealed by the HCA are sufficiently complete that we can use it to deduce the structure of the model's Bayesian networks,
\emph{i.e.}, any conditional dependencies induced by the end-of-inflation hypersurface are negligible in comparison to other dependencies at higher energies.  This assumption will not be valid in general, but we leave the detailed analysis of more complicated potentials for future work.

\paragraph{Isocurvature, reheating, and $N_*$.---}
We compute the power spectrum at the end of inflation, when the evolution of the universe may be non-adiabatic.  The field perturbations at the end of inflation may still have relative entropy perturbations that yield a non-adiabatic pressure perturbation $\delta P_\mathrm{nad}$, which sources a change in the super-horizon curvature perturbation~\cite{GarciaBellido:1995qq,Wands:2000dp,Malik:2002jb}
\begin{equation}
  \dot \zeta = - \frac{H}{\sum_i \dot \phi_i^2} \delta P_\mathrm{nad},
  \label{eqn:XXX}
\end{equation}
where an overdot indicates a derivative with respect to cosmic time.\footnote{We have also assumed that $\zeta \approx \mathcal R$ when $k \gg aH$, where $\mathcal R$ is the curvature perturbation on comoving hypersurfaces.}  Consequently, the primordial spectrum $P_\zeta (k)$ can continue to evolve until it reaches the adiabatic limit, where $\delta P_\mathrm{nad} \to 0$.
We  showed in Ref.~\cite{Easther:2013rva} that the non-adiabatic power spectrum increases with the number of fields $N_f$ for the potential~\eqref{eqn:Nquad}, although what this means for reheating dependence is still unknown.  However, Refs.~\cite{Leung:2012ve,Huston:2013kgl,Leung:2013rza,Meyers:2013gua} demonstrate that for two fields perturbative reheating may yield a significant change in the predicted value of $n_s$, $r$, and $f_\mathrm{NL}$, if there are significant isocurvature perturbations at the end of inflation.  Similarly, for some masses this model can realize a curvaton-like mechanism that may suppress $r \sim 0$ after inflation ends~\cite{Hardwick:2015tma,Vennin:2015vfa}. Consequently, the predicted distributions for $n_s$, $\alpha_s$, and $r$ found in Sects.~\ref{ssect:pheno}--\ref{ssect:hep} do not include  super-horizon effects from reheating and may thus be altered by the largely uncertain physics of the post-inflationary universe.

In principle, $N_*$ also depends on the post-inflationary expansion history by~\cite{Adshead:2010mc,Easther:2011yq}
\begin{align}
  N_* \approx \, & 56.12 + \frac{1}{3 (1+w_r)} \log \left( \frac{2}{3} \right) + \frac{1}{4} \log \left( \frac{V_*}{V_c} \right) \\
  &+ \frac{1-3w_r}{12(1+w_r)} \log \left( \frac{\rho_r}{V_c} \right) + \log \left( \frac{V_*^{\frac{1}{4}}}{10^{16} \, \mathrm{GeV}} \right), \notag
  \label{eqn:XXX}
\end{align}
which is a generalization of the standard matching equation~\cite{Liddle:1993fq,Liddle:2003as} to include an integrated equation of state $w_r$ for the reheating epoch.
The primary uncertainty in $N_*$ comes from the unknown physics of reheating and the primordial dark ages~\cite{Boyle:2005se} prior to neutrino decoupling, which is parametrized by $w_r$ and the energy density $\rho_r$ when the universe is guaranteed to have been thermalized.  Consequently, a prior probability for $N_*$ encodes an expectation on the dynamics of homogeneous background quantities during reheating.  If we assume that $\rho_r^{1/4} \gtrsim \mathcal O(100 \, \mathrm{MeV})$, then a matter-dominated equation of state $w_r=0$ gives $N_* \gtrsim 40$, while a more exotic equation of state $w_r \approx 1$ gives an upper bound of $N_* \lesssim 70$ for $N_f$--quadratic inflation.  Although the lower end of this range is generally favored \emph{a priori}~\cite{Liddle:2003as,Martin:2010kz,Mortonson:2010er,Easther:2011yq}, we set these values as our upper and lower limits for $N_*$.

\subsection{BayesNet A: a simple model}
\label{ssect:pheno}

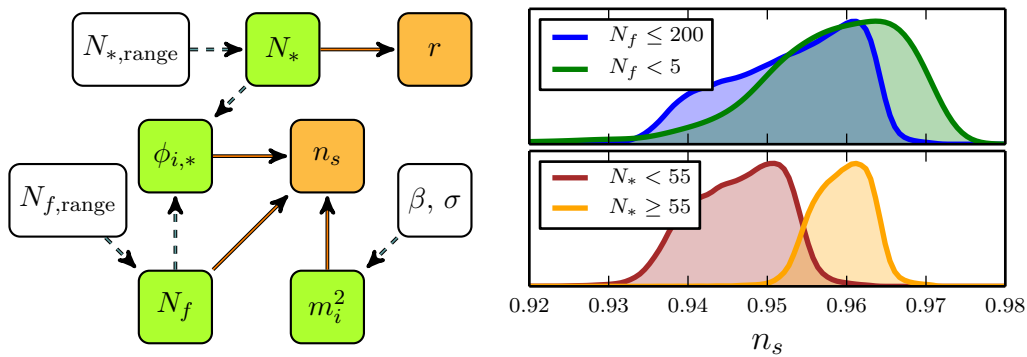
\begin{figure}
  \centering
  \begin{subfigure}[c]{0.4\textwidth}
  \centering
  \begin{tikzpicture}[>=stealth',shorten >=1pt,node distance=2.0cm,on grid,
    initial/.style    ={},
    block/.style={draw=black, thick, rounded corners,
    minimum width=2.5em,
    minimum height=2.5em},
  ]
  \tikzset{every node/.style={fill=white}}
  \node[block]          (ab)     {$N_{*,\mathrm{range}}$};
  \tikzset{every node/.style={fill=GreenYellow}}
    \node[block]          (Npiv) [right = of ab]    {$N_*$};
  \tikzset{every node/.style={fill=Dandelion}}
    \node[block]          (r) [right  = of Npiv]    {$r$};
  \tikzset{every node/.style={fill=GreenYellow}}
    \node[block]          (phipiv) [below left  = of Npiv]    {$\phi_{i,*}$};
  \tikzset{every node/.style={fill=Dandelion}}
    \node[block]          (ns) [right  = of phipiv]    {$n_s$};
  \tikzset{every node/.style={fill=GreenYellow}}
    \node[block]          (Nf) [below  = of phipiv]    {$N_f$};
    \node[block]          (m) [right  = of Nf]    {$m_i^2$};
  \tikzset{every node/.style={fill=white}}
    \node[block]          (beta) [above right  = of m]    {$\beta$, $\sigma$};
    \node[block]          (Nrange) [above left = of Nf]                         {$N_{f,\mathrm{range}}$};
  \tikzset{every node/.style={fill=Dandelion}}
  \tikzset{every node/.style={fill=white}}
  \tikzset{every node/.style={fill=Tan}}

  \tikzset{mystyle/.style={->,double=CadetBlue,dashed}}
  \path (ab)     edge [mystyle]     (Npiv);
  \path (beta)     edge [mystyle]     (m);
  \path (Npiv)     edge [mystyle]     (phipiv);
  \path (Nf)     edge [mystyle]     (phipiv);
  \path (Nrange)     edge [mystyle]     (Nf);
  \tikzset{mystyle/.style={->,double=orange}}
  \path (Nf)     edge [mystyle]     (ns);
  \path (Npiv)     edge [mystyle]     (r);
  \path (phipiv)     edge [mystyle]     (ns);
  \path (m)     edge [mystyle]     (ns);
  \end{tikzpicture}
  \end{subfigure}
  \begin{subfigure}[c]{0.48\textwidth}
  \centering
  \includegraphics{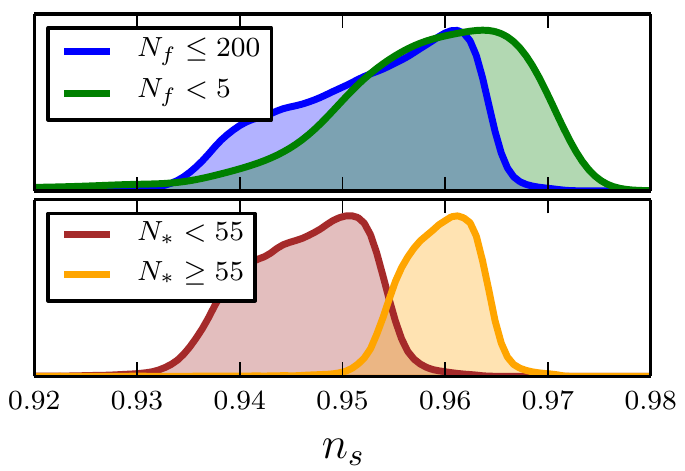}
  \end{subfigure}
  \caption{ \label{fig:DAG_nquad1} (\emph{Left}) Bayesian network (BayesNet A) for $N_f$--quadratic inflation with masses $m_i^2$ distributed by Eq.~\eqref{eqn:MP}.  We fix $\beta=0.5$ and $\sigma=5\e{-6}$ and the number of fields is distributed as $N_f \sim \mathcal U \left[1,200 \right]$.  Any background field values $\phi_{i,*}$ at horizon-crossing ($k_*=a_*H_*$) are equally likely according to the iso-$N_*$ prior.  We draw $N_*$ uniformly from $N_* \sim \mathcal U[40,70]$.  (\emph{Right}) Marginalized probability distributions $P(n_s)$ with varying ranges for $N_f$ and $N_*$.  The marginalized prediction is sensitively dependent only on the upper limit allowed for $N_*$. }
\end{figure}

\paragraph{Model.---}
Figure~\ref{fig:DAG_nquad1} displays the Bayesian network for our first version of $N_f$--quadratic inflation, a phenomenological description of assisted inflation around a minimum in a high dimensional field space. We use an uninformative, uniform prior probability for the number of $e$-folds of expansion $N_* \sim \mathcal U [40,70]$.
We set $N_f < 200$, which does not strongly affect the model's predictions, as we will discuss below, and choose a uniform prior probability over this range, $N_f \sim \mathcal U[1,200]$.  In Sect.~\ref{ssect:maxent} we update this prior to a maximum entropy prior.

We assume that the fields are slowly rolling, and the initial velocities $\dot{\phi}_{i,0}$ are those predicted by the slow-roll equations of motion~\eqref{eqn:kg}.
Following Refs.~\cite{Frazer:2013zoa,Easther:2013rva,Price:2014ufa}, we set the initial field values $\phi_{i,0}$ with the \emph{iso-$N_*$ prior},  a uniform probability distribution $f_{\mathrm{iso}N_*}$ pulled back onto an $N_f$--dimensional hypersphere in field-space,  
\begin{equation}
  f_{\mathrm{iso}N_*}(\phi_{i,0} \| N_f) \propto \delta \left( \sum_i^{N_f} \phi_{i,0}^2 - r_{\mathrm{iso}N_*} \right),
  \label{eqn:f_isoN}
\end{equation}
where $\delta$ is the Dirac delta function and $r_{\mathrm{iso}N_*}$ is the radius of the hypersphere.
We set
  $r_{\mathrm{iso}N_*} = 2 \sqrt{N_*}$
since, in the slow-roll limit, the number of $e$-folds between the initial condition and the end of inflation is approximately
  $N_\mathrm{tot}  \approx \frac{1}{4} \sum_i \phi_{i,0}^2$
if we ignore the contribution to $N_\mathrm{tot}$ from $\phi_{i,c}$.
Then $f_{\mathrm{iso}N_*}$ is defined on surfaces of constant $e$-foldings and by fixing $N_\mathrm{tot}=N_*$ we can identify the initial field values with those at horizon crossing $\phi_{i,0}=\phi_{i,*}$.  This reduces the  number of parameters in the model and allows efficient sampling of initial conditions, since sampling Eq.~\eqref{eqn:f_isoN} amounts to picking  uniformly distributed points on an $(N_{f}-1)$-sphere.  The specific choice of prior for the initial conditions has a very weak impact on the predictions for the primordial power spectra in this model~\cite{Easther:2013rva}.

Following Ref.~\cite{Easther:2005zr} we assume that the masses $m_i^2$ are the eigenvalues of a random matrix
\begin{equation}
  \mathcal M_{ij}^2 \sim A_{ik} A^{k}_{\; j},
  \label{eqn:mass_matrix}
\end{equation}
where $A_{ij}$ is a real matrix with independent and identically distributed (iid) random entries that have a distribution with mean $\mu$ and standard deviation $\sigma$.  We assume $\mu=0$ and choose $\sigma=5\e{-6}$ to obtain a primordial scalar spectrum that has amplitude $A_s \sim 10^{-9}\mathrm{-}10^{-10}$.  For convenience we fix $\sigma$, since it primarily affects only $A_s$, as demonstrated in Ref.~\cite{Easther:2005zr}.  If the number of fields is large, the distribution for the eigenvalues of $\mathcal M_{ij}^2$ limits toward the Mar\v{c}enko-Pastur distribution with hyperparameters $\beta$ and $\sigma$, defined by
\begin{equation}
  P_\mathrm{MP} (m^2 \, | \, \sigma, \beta) = \frac{1}{2 \pi m^2 \beta \sigma^2} \sqrt{ \left(\beta_+ - m^2 \right) \left(m^2 - \beta_- \right) },
  \label{eqn:MP}
\end{equation}
where $\beta_\pm \equiv \sigma^2 (1 \pm \sqrt{\beta})^2$ and $\beta=\#^A_\mathrm{row}/\#^A_\mathrm{colm}$ is the ratio of the number of rows to the number of columns in the matrix $A_{ij}$ in Eq.~\eqref{eqn:mass_matrix}.  We set $\beta=0.5$, which is motivated by the expected ratio of axions to the total number of moduli in the stringy construction of Ref.~\cite{Easther:2005zr}.
This choice sets the typical mass range so that all of the masses are within an order of magnitude of each other.

\begin{figure}
\centering
\includegraphics[width=1.0\textwidth]{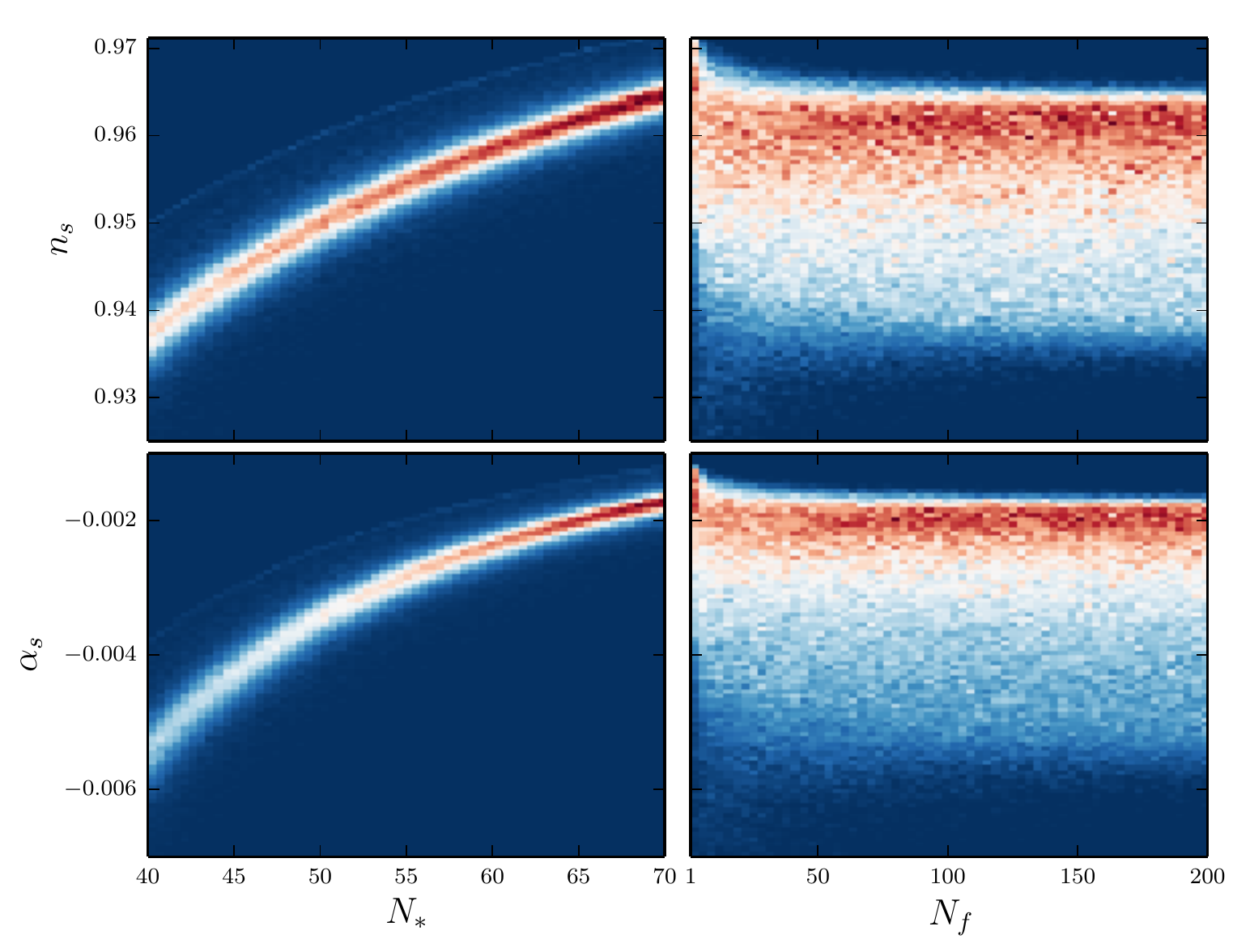}
\caption{\label{fig:model1_hist}  Histogram estimation for the marginalized distributions (\emph{upper left}) $P(n_s \| N_*)$, (\emph{upper right}) $P(n_s \| N_f)$, (\emph{lower left}) $P(\alpha_s \| N_*)$, and (\emph{lower right}) $P(\alpha_s \| N_f)$ of BayesNet A.  Each vertical column of bins is independently normalized.  The prediction for $r$ is given by Eq.~\eqref{eqn:Pr}.}  The marginalized distributions for $n_s$ and $\alpha_s$ change only minimally for $N_f \gtrsim 10$ but depends strongly on the choice of $N_*$.
\end{figure}

\paragraph{Predictions.---}

The marginalized probability distribution for the tensor:scalar ratio $r$ in the HCA can be obtained analytically by integrating Eq.~\eqref{eqn:Nquad_r} over $N_*$ with the prior $N\sim \mathcal U[a,b]$, giving
\begin{equation}
  P(r \| a, b ) =
    \begin{cases}
      \left[8/(b-a)\right] \; r^{-2} & 8/b \leq r \leq 8/a \\
      0 & \mathrm{else}
    \end{cases}.
  \label{eqn:Pr}
\end{equation}
This distribution maximizes at $r=8/b$, which is $r = 0.114$ for the upper limit on $N_*\le 70$ for this model.  Consequently, when using this generative model to predict the tensor-to-scalar ratio there will be a strong dependence on the upper limit that we allow for $N_*$.
In practice we do not use the HCA, but solve Eq.~\eqref{eqn:r_sr} directly for $r$ to obtain correct results to first order in the slow-roll parameters, but the two approaches are in good agreement, consistently with Ref.~\cite{Easther:2013rva}.  The generic prediction of $r \sim 0.1$ holds for any reasonable range of allowed values for $N_*$.  Unless $r$ is modified by reheating effects not incorporated in this analysis, this value is in tension with the latest results from the joint \emph{Planck} and BICEP2/\emph{Keck Array}~\cite{Ade:2013zuv,Ade:2013rta,Mortonson:2014bja,Flauger:2014qra,Ade:2015tva,Ade:2015xua}.  Here, $N_f$--quadratic inflation was chosen only as an example model to display the Bayesian hierarchical modelling techniques, but in a different inflation scenario with $r \lesssim 0.1$ the exact prediction for $r$ and its relative dependencies on the hyperparameters would be highly important.

Figures~\ref{fig:DAG_nquad1}~and~\ref{fig:model1_hist} show the more complicated predictions for the spectral tilt $n_s$ and scalar running $\alpha_s$.  Figure~\ref{fig:DAG_nquad1} plots $P(n_s)$ as a function of the hyperparameters that control the ranges of $N_f$ and $N_*$.  When the upper limit on the number of fields is between 20 and 200  the predicted values for $n_s$ or $\alpha_s$ do not change.  For example, the distribution
\begin{equation}
  P(n_s \| N_f < 20) \equiv \sum_{i=1}^{20} P(n_s \| N_f = i) \, P(N_f = i)
  \label{eqn:XXX}
\end{equation}
closely matches the plotted curve $P(n_s \| N_f < 200)$ in Fig.~\ref{fig:DAG_nquad1}.
However, if the upper limit on the number of fields is less than ten, the marginal distribution $P(n_s)$ can depart substantially from the distribution $P(n_s \| N_f < 200)$.
Figure~\ref{fig:model1_hist} plots the conditional probabilities $P(n_s \| N_f)$ and $P(\alpha_s \| N_f)$, demonstrating that the distributions quickly approach a limiting distribution
\begin{equation}
  P_\infty(n_s) \equiv P(n_s \| N_f \to \infty) \qquad \mathrm{and} \qquad P_\infty(\alpha_s) \equiv P(\alpha_s \| N_f \to \infty)
  \label{eqn:XXX}
\end{equation}
for $N_f \gtrsim 20$.
Consequently, the marginalized distributions $P(n_s)$ and $P(\alpha_s)$ will be dominated by the contributions from the large-$N_f$ region of parameter space:
\begin{equation}
  P(n_s) \sim \int P(n_s \| N_f) \; dN_f \sim P_\infty(n_s)
  \label{eqn:XXX}
\end{equation}
and
\begin{equation}
  P(\alpha_s) \sim P_\infty(\alpha_s),
  \label{eqn:XXX}
\end{equation}
assuming that the upper limit for $N_f$ is substantially larger than $N_f \gtrsim 20$.

\begin{figure}
\centering
\includegraphics[width=1.0\textwidth]{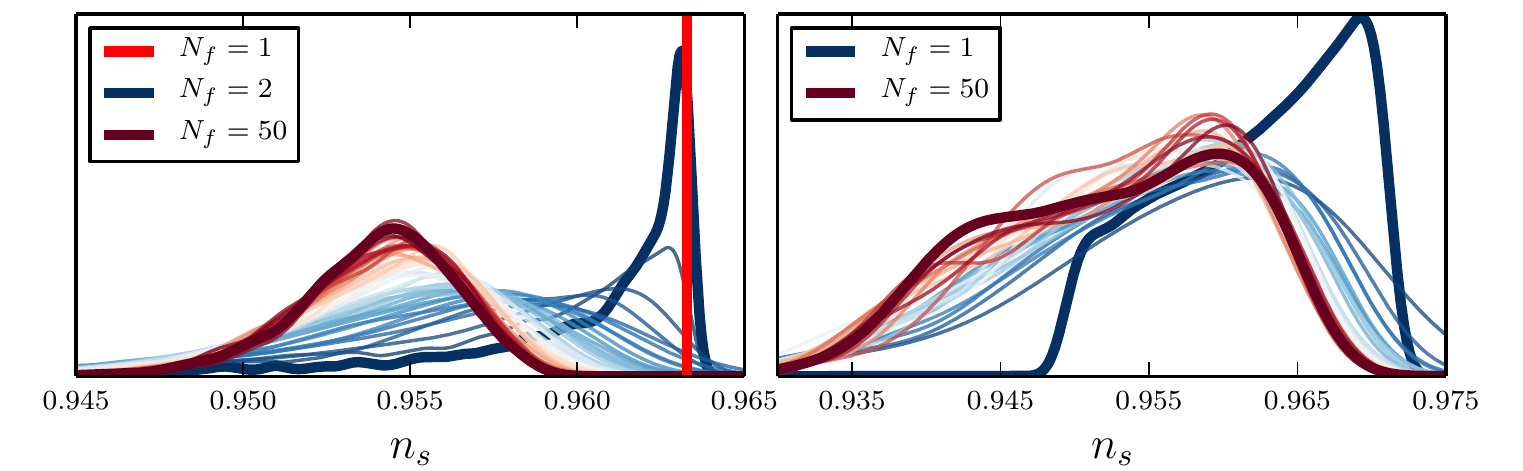}
\caption{\label{fig:model1_ns_transition} (\emph{Left}) The probability distribution $P(n_s \| N_f, N_*=55)$ with $1 \le N_f \le 50$ and (\emph{Right}) the probability distribution $P(n_s \| N_f)$ with $N_*$ marginalized over a uniform distribution $\mathcal U [40,70]$ as in Fig.~\ref{fig:model1_hist} for BayesNet A.}
\end{figure}

Figure~\ref{fig:model1_ns_transition} illustrates the transition to a limiting distribution as $N_f$ is increased, with $N_*$ both fixed and as a free parameter.  The distribution $P(n_s \| N_f=1, N_*=55)$ is single-valued at $n_s = 0.963$, matching $m^2 \phi^2$ inflation.  As $N_f$ is increased the peak of the distribution remains near $n_s = 0.963$ with an increasing variance for $N_f \lesssim 5$.  For larger values of $N_f$ the distribution approaches a normal distribution with mean of $n_s =0.954$, as detailed in Ref.~\cite{Easther:2013rva}.  The central limit theorem suggests that the variance of the distribution for $n_s$ will be inversely proportional to $N_f$.  However, when $N_*$ is marginalized, the transition to the limiting behavior $P_\infty(n_s)$ is very rapid.  For single-field inflation, the predicted spectral tilt is $n_s \sim N_*^{-1}$ which can be analytically marginalized, as shown in Eq.~\eqref{eqn:Pr}.  However, for more fields the distribution $P(n_s \| N_f)$ quickly approaches a limiting case, which does not substantially differ over the range $2 < N_f< 50$ that is plotted in Fig.~\ref{fig:model1_ns_transition}.
We do not expect the variance of this distribution to vanish in the $N_f \to \infty$ limit due to the inherent uncertainty in $N_*$ that is being integrated over in $P(n_s)$.  Therefore, we can estimate the shape of $P_\infty(n_s)$ by the limiting curve in Fig.~\ref{fig:model1_ns_transition}.

While they are largely independent of the total number of fields, $P(n_s)$ and $P(\alpha_s)$ depend strongly on $N_*$.  Figure~\ref{fig:DAG_nquad1} gives the predicted spectral tilt when the upper and lower limits for the allowed value of $N_*$ are changed.  If $N_*$ is uniformly distributed in the range $55 \le N_* \le 70$, the mean prediction is $\vev{n_s} = 0.959$ with a standard deviation of $\sigma=3.89\e{-3}$.
However, if $N_*$ is uniformly distributed in the range $40 \le N_* < 55$, then $\vev{n_s}=0.946$ and the standard deviation grows to $\sigma=6.03\e{-3}$.  As in Eq.~\eqref{eqn:Pr} for the distribution of $r$, $P(n_s)$ is most strongly dependent on the upper limit allowed for $N_*$.
The left and right columns of Fig.~\ref{fig:model1_hist} show that for all values of $N_*$ the variance of the distributions $P(n_s \| N_*)$ and $P(\alpha_s\| N_*)$, which marginalizes over $N_f$, is significantly less than $P(n_s \| N_f)$ and $P(\alpha_s\| N_f)$, which marginalizes over $N_*$.
While this demonstrates qualitatively that $N_*$ is the  most important parameter for the prediction of $n_s$ and $\alpha_s$, not all observables will be sensitively dependent on $N_*$ in this model, \emph{e.g.}, the prediction for the tensor consistency relation $P(n_t/r)$ has a variance that scales like $\sigma^2 \sim 1/N_f$ even after marginalizing over $N_*$~\cite{Price:2014ufa}.

\paragraph{The effectiveness of the HCA.---}
In the horizon crossing approximation (HCA) we ignore the contribution of the field values at the end of inflation to the total number of $e$-folds of expansion during inflation, as in Eq.~\eqref{eqn:Ntot}.  While this approximation holds for $\phi_{i,*} \gg \phi_{i,c}$, Fig.~\ref{fig:field_evol} shows that $N_f$--quadratic inflation has many
light fields $\phi_i$ with small masses, which evolve only minimally during inflation.
However, even though this model does not always have a large hierarchy between the horizon crossing and end-of-inflation field values, the marginalized predictions obtained in this Section are still closely approximated by the HCA.  We interpret this as a consequence of the relative unimportance of the parameter $N_f$ in determining the primordial power spectrum in this model.

The model's predictions depend on the rate of change of the number of $e$-folds as a function of the horizon crossing field values~\cite{Battefeld:2006sz}, which is given by
\begin{align}
  \label{eqn:dN_BE}
dN = & \left (\frac{\partial N}{\partial \phi_{i,*}} + \frac{\partial N}{\partial \phi_{j,c}} \frac{\partial \phi_{j,c}}{\partial \phi_{i,*}} \right) d \phi_{i,*} \\
  = & \sum_{j=1}^{N_f} \left[ \left( \frac{V_j}{V_j'} \right)_* - \sum_i \frac{\partial \phi_{i,c}}{\partial \phi_{j,*}} \left(\frac{V_j}{V_j'} \right)_c \right] d \phi_{j,*}. \notag
\end{align}
The HCA is equivalent to ignoring the second term in the brackets.
To zeroth order the field values for the lightest fields at horizon crossing are approximately the same as at the end of inflation, which yields
\begin{equation}
  \frac{\partial \phi_{i,c}}{\partial \phi_{j,*}}  \approx \delta_{ij}.
  \label{eqn:dphi}
\end{equation}
Substituting Eq.~\eqref{eqn:dphi} into Eq.~\eqref{eqn:dN_BE} reduces the range of the summation so that only the fields that evolve substantially during inflation contribute to the derivatives of $N$, which reduces the effective $N_f$ for this model.  Since Fig.~\ref{fig:model1_hist} shows that the marginalized predictions for the spectral tilt and scalar running reach a limiting distribution for $N_f \gtrsim 20$, this moderate reduction of $N_f$ will not typically affect the prediction at the end of inflation.  Since the HCA is a good estimate for those fields with large field displacements during inflation and since removing the light fields from the summation does not change the prediction, the HCA should therefore closely resemble the exact solution for this model even when $\phi_{i,*} \sim \phi_{i,c}$.

\subsection{BayesNet B: maximum entropy prior for $N_f$}
\label{ssect:maxent}

\paragraph{Model.---}
In BayesNet A of Sect.~\ref{ssect:pheno} the predictions for $n_s$ and $\alpha_s$ do not depend sensitively on the number of fields and quickly approaches a limiting value for $N_f \gtrsim 20$.  However, if the prior probability for the number of fields were weighted  toward $N_f \sim 1$ this parameter may have more impact on the model's predictions.  Since we had no fundamental reason for choosing the uniform distribution $N_f \sim \mathcal U [1,200]$ in Sect.~\ref{ssect:pheno}, we now assess how the results of the previous model depend on the chosen prior distribution  for $N_f$.

The resulting Bayesian network is similar to that of Fig.~\ref{fig:DAG_nquad1}, but uses a maximum entropy prior  for $N_f$, \emph{i.e.}, a prior probability that contains the minimal amount of Shannon information~\cite{shannon2001mathematical}.
We allow  $N_f = \left\{1,2,\dots \right\}$ and fix the expected size of $N_f$ \emph{a priori}, which is necessary if we are to define a maximum entropy distribution over the positive integers.  The geometric prior probability $\mathcal G_p(N_f)$ is the probability mass function that maximizes the Shannon entropy for this range of allowed values, 
\begin{equation}
  P\left(N_f \| \vev{N_f} \right) \equiv \mathcal G_p(N_f) = p \; (1-p)^{N_f-1}
  \label{eqn:geom}
\end{equation}
for $p \equiv 1/\vev{N_f}$.  The mean and variance of $\mathcal G_p$ is $p$ and $(1-p)/p^2$, respectively.  The distribution peaks at $N_f=1$ for all $p$, giving a  larger weighting toward lower numbers of fields relative to a uniform distribution.
We then add the \emph{a priori} expected number of fields $\vev{N_f}$ to the Bayesian network of Fig.~\ref{fig:DAG_nquad1} as a hyperparameter for $N_f$.  Although this seems like additional structure that we have added to BayesNet A, the principle of maximum entropy ensures that our arbitrary choice of boundary for the uniform prior $N_f \sim \mathcal U[1,200]$ contains more assumptions than Eq.~\eqref{eqn:geom} when measured according to the Shannon information.

\begin{figure}
\centering
\includegraphics[width=1.0\textwidth]{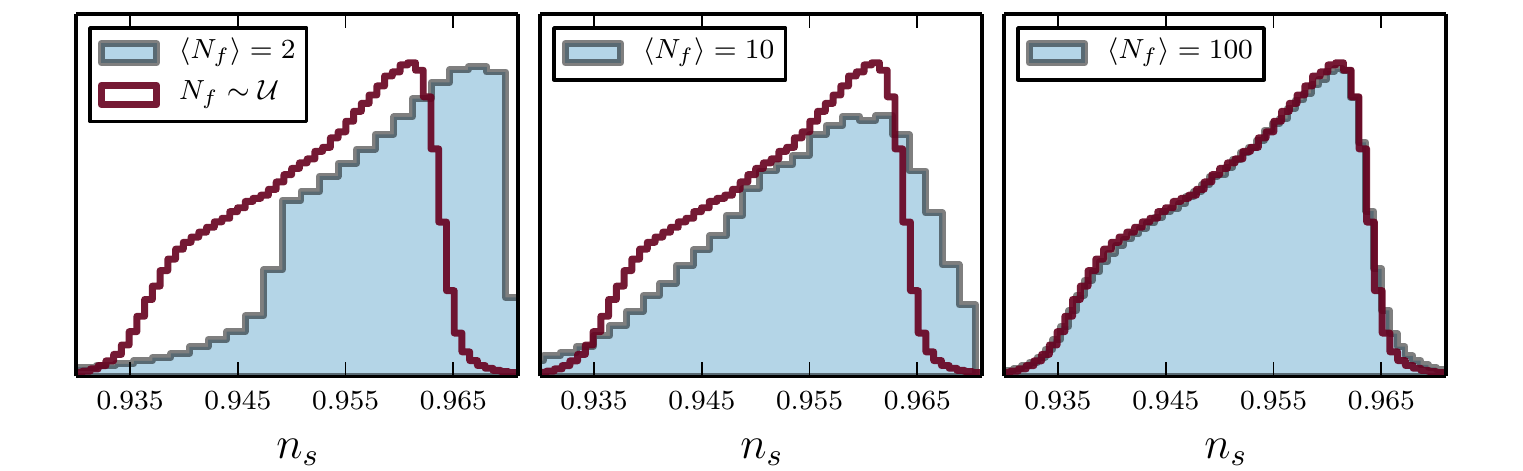}
\caption{\label{fig:model3_hist}  Histogram estimation of $P(n_s \| \vev{N_f})$ for BayesNet B with $N_f \sim \mathcal G_{p}$ in Eq.~\eqref{eqn:geom} as compared to the prediction from the uniform prior $N_f \sim \mathcal U[1,200]$ of BayesNet A.}
\end{figure}

\paragraph{Predictions.---}
Figure~\ref{fig:model3_hist} compares the marginalized prediction $P(n_s \| \vev{N_f})$ for three different choices of $\vev{N_f}$ to the uniformly distributed case of Sect.~\ref{ssect:pheno}.  We plot distributions for a model with few expected fields $\vev{N_f}=2$, an intermediate number of fields $\vev{N_f}=10$, and many fields $\vev{N_f}=100$, where the specific numbers were chosen based on the behavior of $P(n_s \| N_f)$ in Fig.~\ref{fig:model1_hist}.  If we pick $\vev{N_f}=1$, then there is zero probability for $N_f>1$ and we recover the single-field inflation results.
If we expect many fields $\vev{N_f}=100$, then we obtain the same limiting distribution $P_\infty (n_s)$ that we saw in Fig.~\ref{fig:model1_ns_transition}.  Out of a sample of $975,000$ points the largest simulated value was $N_f=983$ and $50\%$ of the sample had $N_f <70$.  The mean predicted value of the spectral index is $\vev{n_s}=0.953$, with a standard deviation of $\sigma=8.73\e{-3}$.   If we expect a large number of fields, then the marginalized prediction will be largely independent of the exact value of $N_f$.

As we expect from $P(n_s \| N_f)$ in Fig.~\ref{fig:model1_hist}, for lower $\vev{N_f}$ the scalar spectrum is marginally more blue.  However, it is not well approximated by the single-field prediction, which has a mean value of $\vev{n_s}=0.962$, due to the sharp transition in the functional form of $P(n_s \| N_f)$ when the number of fields changes from $N_f=1$ to $N_f=2$, as displayed in Fig.~\ref{fig:model1_ns_transition}.
For $\vev{N_f}=2$ the vast majority of samples have $N_f \le 2$ and the largest simulated value was $N_f=20$ in a $199,000$ element sample.  Approximately $50\%$ of this sample has $N_f>1$, which results in a redder spectrum than if $N_f$ were fixed to unity, with a mean prediction of $\vev{n_s}=0.958$ and a standard deviation of $\sigma=1.47\e{-2}$.
For $\vev{N_f}=10$ the distribution for $n_s$ is intermediate between these two cases, giving $\vev{n_s}=0.955$ and $\sigma=1.29\e{-2}$.
Consequently, if we expect $\vev{N_f} \gtrsim 10$, then the predictions for $n_s$ from the maximum entropy prior for $N_f$ should be well approximated by the results of Sect.~\ref{ssect:pheno}.

\subsection{BayesNet C: high-energy dependence}
\label{ssect:hep}

\begin{figure}
  \centering
  \begin{subfigure}{0.48\textwidth}
  \begin{tikzpicture}[>=stealth',shorten >=1pt,node distance=2.0cm,on grid,
    initial/.style    ={},
    block/.style={draw=black, thick, rounded corners,
    minimum width=2.5em,
    minimum height=2.5em},
  ]
  \tikzset{every node/.style={fill=Dandelion}}
    \node[block]          (r)                         {$r$};
  \tikzset{every node/.style={fill=GreenYellow}}
  \node[block]          (Npiv) [left = of r]                         {$N_*$};
  \tikzset{every node/.style={fill=white}}
  \node[block]          (ab) [left = of Npiv]                         {$N_{*,\mathrm{range}}$};
  \tikzset{every node/.style={fill=GreenYellow}}
  \node[block]          (phipiv) [below = of r]                         {$\phi_{i,*}$};
  \tikzset{every node/.style={fill=white}}
  \node[block]          (phi0) [left = of phipiv]                         {$\phi_{i,0}$};
  \tikzset{every node/.style={fill=GreenYellow}}
  \node[block]          (m) [below = of phi0]                        {$m_i^2$};
  \tikzset{every node/.style={fill=white}}
  \node[block]          (beta) [below left  of = m, node distance = 3.0 cm]                         {$\beta$, $\sigma$};
  \tikzset{every node/.style={fill=Tan}}
  \node[block]          (betarange) [right = of beta]                         {$\beta_\mathrm{range}$};
  \tikzset{every node/.style={fill=Dandelion}}
  \node[block]          (ns) [below  of = phipiv, node distance = 5cm]                        {$n_s$};
  \tikzset{every node/.style={fill=GreenYellow}}
  \node[block]          (Nf) [left  of = ns, node distance = 6cm]                        {$N_f$};
  \tikzset{every node/.style={fill=Tan}}
  \node[block]          (phimax) [above  of = Nf, node distance = 5.0cm]  {$\phi_{\mathrm{max}}$};
  \tikzset{every node/.style={fill=white}}
  \node[block]          (Nrange) [below right of = phimax]                         {$N_{f,\mathrm{range}}$};
  \tikzset{every node/.style={fill=Tan}}
  \node[block]          (rho0) [above  of = phimax, node distance=1.5cm]                         {$\rho_0$};

  \tikzset{mystyle/.style={->,double=orange}}
  \path (Npiv)     edge [mystyle]     (r);
  \path (phipiv)     edge [mystyle]     (ns);
  \path (m)     edge [mystyle]     (ns);
  \path (Nf)     edge [mystyle]     (ns);
  \path (Npiv)     edge [mystyle]     (phipiv);
  \path (phi0)     edge [mystyle]     (phipiv);
  \path (m)     edge [mystyle]     (phipiv);

  \tikzset{mystyle/.style={->,double=CadetBlue,dashed}}
  \path (ab)     edge [mystyle]     (Npiv);
  \path (rho0)     edge [mystyle]     (phi0);
  \path (m)     edge [mystyle]     (phi0);
  \path (Nf)     edge [mystyle]     (phi0);
  \path (Nrange)     edge [mystyle]     (Nf);
  \path (beta)     edge [mystyle]     (m);
  \path (betarange)     edge [mystyle]     (beta);
  \path (phimax)     edge [mystyle]     (phi0);
  \path (phimax)     edge [mystyle]     (Nf);

  \end{tikzpicture}
  \end{subfigure}
  \begin{subfigure}[c]{0.48\textwidth}
  \centering
  \includegraphics{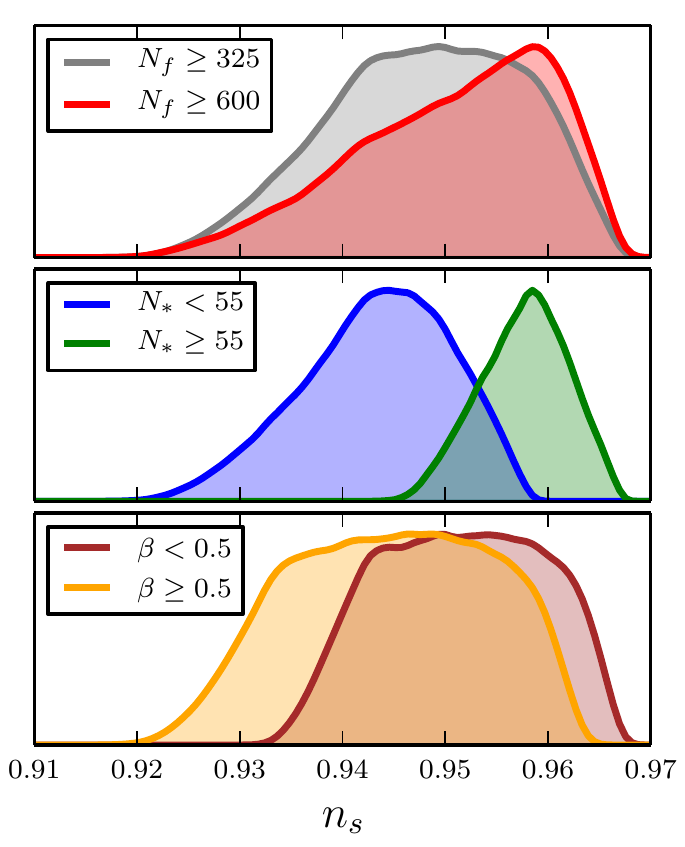}
  \end{subfigure}
  \begin{subfigure}[c]{1.0\textwidth}
  \centering
  \includegraphics{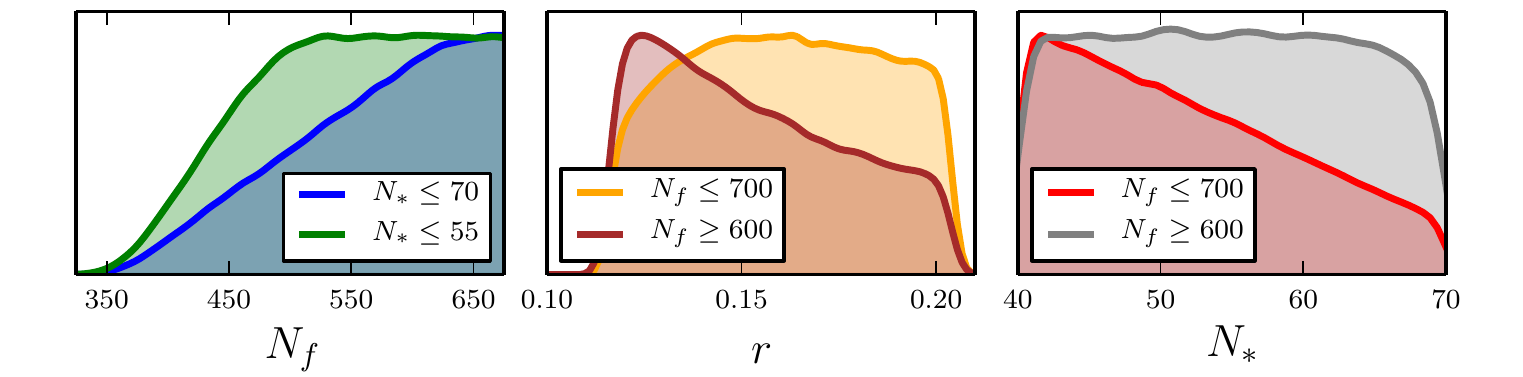}
  \end{subfigure}
  \caption{(\emph{Top-left}) BayesNet C with initial energy $\rho_0= \mpl$ and restrictions $|\phi_{i,0}| < \mpl \equiv \phi_{\max}$ and $N_\mathrm{tot}>N_*$.
    The priors are $m_i^2\sim$ Eq.~\eqref{eqn:MP}, $\sigma=5\e{6}$, and $\beta \sim \mathcal U[0.25,0.75]$; we also sample $N_f \sim \mathcal U[200, 700]$, $\phi_{i,0}(\rho_0) \sim \mathrm{iso-}E$, and $N_* \sim \mathcal U[40,70]$, but enforce the restriction $N_\mathrm{tot} > N_*$ (Eq.~\eqref{eqn:enough_infl}) after $N_f$, $\phi_{i,0}$, and $N_*$ are sampled, which results in non-standard priors for these parameters, which we plot in Fig.~\ref{fig:priors_rejsample}.
(\emph{Top-right})  Marginalized distributions for $n_s$ with varying hyperparameters.
(\emph{Bottom}) Marginalized prediction for the tensor-to-scalar ratio $r$ and the non-uniform priors $P(N_f)$ and $P(N_*)$, with varying hyperparameter ranges.  }
  \label{fig:DAG_nquad2}
\end{figure}

\paragraph{Model.---}

In Sects.~\ref{ssect:pheno}~and~\ref{ssect:maxent} we restricted the dimensionality of parameter space by equating the initial conditions $\phi_{i,0}$ and horizon crossing values $\phi_{i,*}$ by using the iso-$N_*$ prior, as well as fixing the hyperparameter $\beta=0.5$ in the mass distribution of Eq.~\eqref{eqn:MP}.  Here we  relax some of these assumptions and demonstrate how the predictions of the inflation model change when a dependence on an initial energy density $\rho_0$ is incorporated. 

Figure~\ref{fig:DAG_nquad1} is a Bayesian network for a model that involves a higher number of parameters and conditional dependencies for $N_f$--quadratic inflation.
We use the same uniform prior $N_* \sim \mathcal U[40,70]$ as in Sect.~\ref{ssect:pheno} and draw masses from the Mar\v{c}enko-Pastur distribution~\eqref{eqn:MP}, but allow the hyperparameter $\beta$ to vary uniformly in the range $\beta \sim \mathcal U[0.25,0.75]$.  We impose a uniform prior on $N_f$, with the range $N_f \sim \mathcal U[1,700]$, where the upper bound is chosen to ensure that the numerical calculations remain tractable.
We set the background initial conditions $\phi_{i,0}$ and $\dot \phi_{i,0}$ according to the \emph{iso-$E$} prior from Refs.~\cite{Easther:2013bga,Easther:2013rva,Easther:2014zga}.  This is a uniform probability distribution defined on a surface of constant energy density $\rho_0$, which is defined as
\begin{equation}
  f_{\mathrm{iso}E}(\phi_{i,0}, \dot \phi_{i,0}) \propto \delta \left(\frac{1}{2} \sum_i^{N_f} \left[ \dot \phi_{i,0}^2 + m_i^2 \phi_{i,0}^2  \right] - \rho_0 \right),
  \label{eqn:isoE}
\end{equation}
where $\delta$ is the Dirac delta function.
We fix the initial energy scale via the conservative choice $\rho_0 = \mpl$, since at this scale new physics is almost guaranteed to appear.  Refs.~\cite{Corichi:2013kua,Sloan:2014jra} have a similar prior for initial conditions defined on surfaces of constant $\rho$, finding that the predictions resulting from this prior are independent of the relatively arbitrary choice of $\rho_0$.  However, Ref.~\cite{Easther:2013rva}  demonstrated that the exact choice of initial conditions prior has a relatively minor effect on the predicted power spectrum, so we will assume Eq.~\eqref{eqn:isoE} sufficiently captures the initial conditions dependence.

To preserve the interpretation of the $N_f$--quadratic potential as  a Taylor expansion around a local minimum we need to specify the domain of validity of the expansion, requiring
\begin{equation}
  | \phi_{i,0}| \le \phi_\mathrm{max} \equiv \mpl.
  \label{eqn:subplanck}
\end{equation}
The basis in which the domain of validity is given by Eq.~\eqref{eqn:subplanck} is not in general the same as the basis  which diagonalises the mass matrix~\cite{Bachlechner:2014hsa,Bachlechner:2014gfa}, but we will assume here that the two bases are the same.

Since  $\phi_{i,0} \ne \phi_{i,*}$ we also require that the total number of $e$-folds $N_\mathrm{tot}$ between the initial state and the end of inflation exceeds $N_*$.  In the slow-roll limit and ignoring  contributions from the end of inflation, Eq.~\eqref{eqn:Ntot} gives the joint constraint
\begin{equation}
  \frac{1}{4} \sum_i^{N_f} \phi_{i,0}^2 \gtrsim N_*,
  \label{eqn:enough_infl}
\end{equation}
which implies that $N_f \gg 4 N_*$ given the maximum value of $\phi_{i,0}<1$.
These conditions were not needed in Sects.~\ref{ssect:pheno}~and~\ref{ssect:maxent} as the models allowed super-Planckian field displacements.

\begin{figure}
\centering
\includegraphics[width=1.0\textwidth]{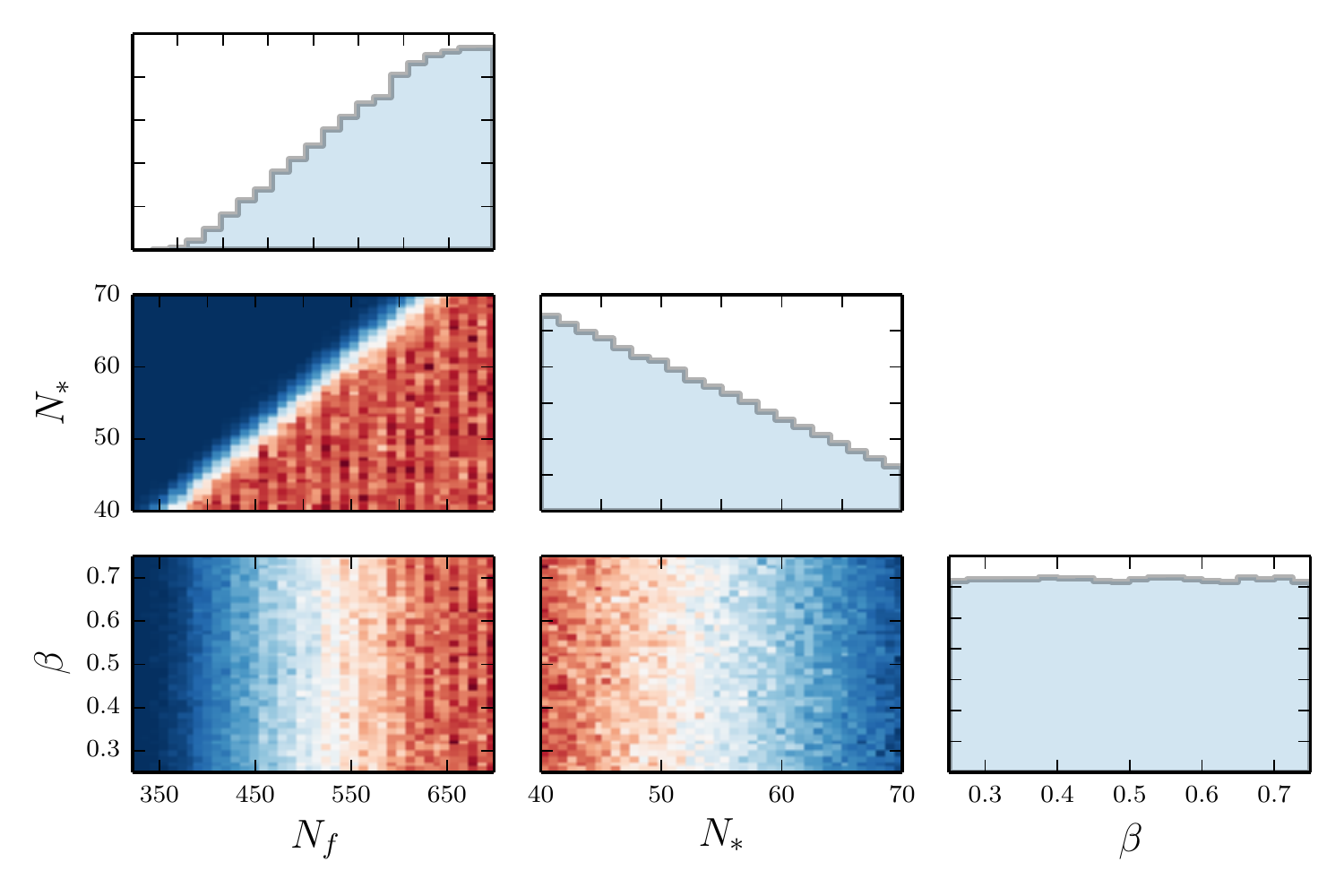}
\caption{Histogram-estimated prior probability distributions for BayesNet C, after enforcing the constraint~\eqref{eqn:enough_infl} by rejection sampling.  These are not generally uniform, since we require sub-Planckian field displacements~\eqref{eqn:subplanck}.}
\label{fig:priors_rejsample}
\end{figure}

We sample $f_{\mathrm{iso}E}$ subject to the additional constraint~\eqref{eqn:subplanck} using the techniques described in Ref.~\cite{Easther:2013bga} and enforce $N_\mathrm{tot}>N_*$ with rejection sampling, \emph{i.e.}, discarding those samples that do not realize enough $e$-folds of inflation.
Rejection sampling alters the na\"ive priors $N_f \sim \mathcal U$, $N_* \sim \mathcal U$, $m_i^2 \sim P_\mathrm{MP}$, or $\phi_{i,0} \sim f_{\mathrm{iso}E}$.  The deviation from the na\"ive priors depends on the order in which the conditional dependencies in the Bayesian network in Fig.~\ref{fig:DAG_nquad2} are sampled.  Here, we first select $N_f$, then the initial field values, then $N_*$ from their na\"ive priors and throw away all combinations of parameters $\left\{ N_f, N_*, \phi_{i,0}, \dots \right\}$ that would violate Eq.~\eqref{eqn:enough_infl}.  Conversely, after generating $N_f$ and the initial conditions we could have sampled an \emph{a priori} conditional distribution $P(N_* \| N_* \lesssim N_\mathrm{tot})$, where we calculate $N_\mathrm{tot}$ from the equations of motion.  If $N_\mathrm{tot}$ for a given configuration is larger than the minimum allowed value for $N_*$, then $N_f$ preserves a uniform prior, with only $P(N_*)$ changed.  These two priors should give different predictions, since they will place different weightings on $N_*$.  Furthermore, they describe different expectations on the physics of reheating and post-inflationary evolution, since $N_*$ depends on the entire expansion history of the universe.  We will show that the constraints in Eqs~\eqref{eqn:subplanck}--\eqref{eqn:enough_infl} introduce a change in the predictions from this Bayesian network that dominates any differences resulting from the other novel parameters in the model.

The non-uniform prior probability on $N_f$ and $N_*$ that results from our rejection sampling technique can be seen in Figs.~\ref{fig:DAG_nquad2}~and~\ref{fig:priors_rejsample}.
We set the minimum value of $N_*$ at $40$ $e$-folds, which requires a minimum of  $N_f \gtrsim 160$ to get a sufficient number of $e$-folds of inflation, but more typically needs $N_f \gtrsim 350$.
If $N_f$ is relatively small, we need a larger field displacement to ensure $N_\mathrm{tot}>N_*$, which conflicts with the maximum allowed range for the initial field values.  Similarly, if $N_*$ is relatively large, then more fields are required to get enough inflation.  This results in an \emph{a priori} preference for smaller values of $N_*$ and larger values of $N_f$ as compared to the uniform priors.  However, the parameter $\beta$, which sets the variance of the prior for the masses~\eqref{eqn:MP}, is not affected by these considerations, as shown in Fig.~\ref{fig:priors_rejsample}.  In the slow-roll calculation, the total number of $e$-folds does not depend on the masses~\eqref{eqn:Ntot}, which gives $\beta$ a negligible dependence on the non-trivial prior space for $N_*$.

\begin{figure}
\centering
\includegraphics[width=1.0\textwidth]{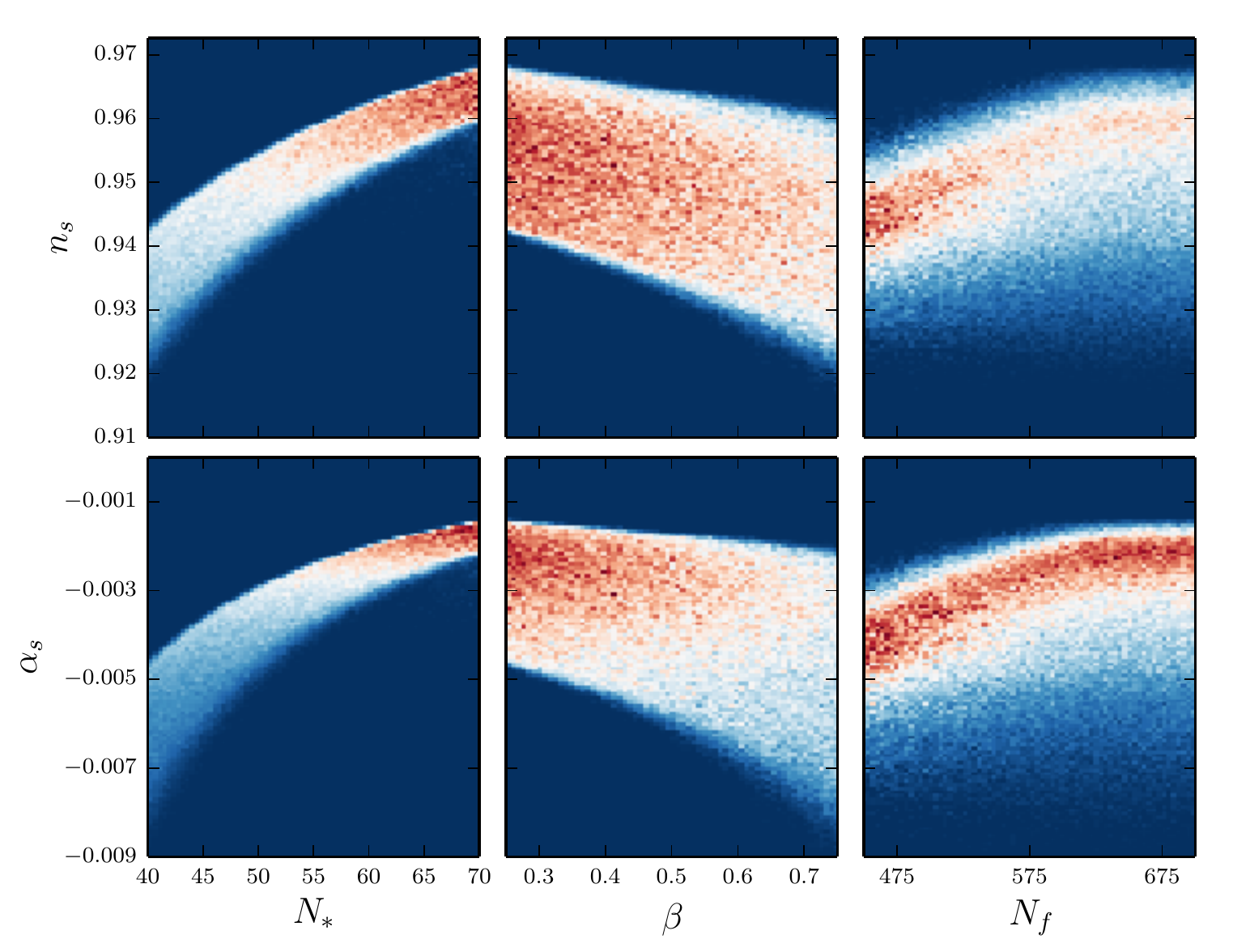}
\caption{\label{fig:model2_hist}  Histogram estimates for the distributions (\emph{left}) $P(n_s \| N_*)$ and $P(\alpha_s \| N_*)$, (\emph{middle}) $P(n_s \| \beta)$ and $P(\alpha_s \| \beta)$, and (\emph{right}) $P(n_s \| N_f)$ and $P(\alpha_s \| N_f)$ for BayesNet C.  Each vertical column of bins is independently normalized as a PDF.  The distributions depend strongly on $N_*$ and $N_f \lesssim 600$ and moderately on $\beta$.  The non-uniform prior probability $P(N_f)$ in Fig.~\ref{fig:DAG_nquad2} makes the predictions moderately dependent on $N_f$.  The prediction for $r$ does not change radically from Fig.~\ref{fig:DAG_nquad2}.}
\end{figure}

\paragraph{Predictions.---}
The prior probability distribution $P(N_*)$ approaches the uniform distribution $\mathcal U[40,70]$ as the number of fields grows large.
Figure~\ref{fig:DAG_nquad2} gives a nearly flat distribution for $N_*$ if we restrict the minimum number of fields to $N_f \ge 600$, which is modified so that it is approximately inversely proportional to $N_*$ when the full range $N_f \le 700$ are included.
The non-uniform prior on $N_*$ gives a marginalized distribution for $r$ which differs from Eq.~\eqref{eqn:Pr}, which is also plotted in Fig.~\ref{fig:DAG_nquad2}.  Since $r \propto N_*^{-1}$ the approximate inverse power law for $P(N_*)$ yields an almost uniform distribution for the tensor-to-scalar ratio in the range $0.11 < r < 0.21$.  This deemphasizes the effect of the upper limit of $N_*$ on the marginalized distribution $P(r)$, which was seen in Eq.~\eqref{eqn:Pr}.
The prior probability for $N_f$ is also given in Fig.~\ref{fig:DAG_nquad2} and begins to approach a flat distribution for smaller values of $N_*$.  For $N_* \le 55$ almost all $N_f  \gtrsim 500$ are almost equally likely, with a decreasing probability for smaller numbers of fields.  In our sample, which is composed of $250,000$ elements, the smallest value was $N_f = 322$, which sets an effective lower bound on the expected number of fields in this model.

Figures~\ref{fig:DAG_nquad2}~and~\ref{fig:model2_hist} display probability distributions for $n_s$ and $\alpha_s$.  The marginalized distributions $P(n_s)$ in Fig.~\ref{fig:DAG_nquad2} are plotted with different allowed ranges of $N_f$, $N_*$, and $\beta$.  The distributions for both $n_s$ and $\alpha_s$ are sensitively dependent on $N_*$ and moderately dependent on $\beta$ and $N_f \lesssim 500$, which can be seen by the relative thickness of the distributions in Fig.~\ref{fig:model2_hist}.  Keeping $\beta$ in the range $0.25 \le \beta \le 0.50$ gives a mean prediction of $\vev{n_s}=0.951$ with standard deviation $\sigma=7.80\e{-3}$, while the range $0.50 \le \beta \le 0.75$ gives $\vev{n_s}=0.945$ and $\sigma=9.50\e{-3}$.  In comparison, if we restrict $55 \le N_* \le 70$, then the mean increases to $\vev{n_s}=0.958$ with standard deviation $\sigma=4.26\e{-3}$, while the range $40 \le N_* \le 55$  gives the redder value of $\vev{n_s}=0.943$ and $\sigma=7.25\e{-3}$.
The dependence on small $N_f$ in the marginalized prediction results from the covariance between $N_f$ and $N_*$, since smaller values of $N_f$ need smaller $N_*$, which gives a redder spectrum as seen in Fig.~\ref{fig:model2_hist}.  The plot for $N_f$ in Fig.~\ref{fig:model2_hist} is truncated at 450 because the lower ranges of $N_f$ are not well sampled and the histogram estimator is noisy.

We see from this analysis  that imposing a simple bound on the field values specified by Eq.~\eqref{eqn:subplanck} has dramatic implications for the model's prediction. Interestingly, although the priors for $N_{*}$ and $N_{f}$ incorporate information about different sectors of fundamental particle theory, this requirement creates a joint \emph{a priori} dependency between these two hyperparameters.\footnote{Note that $N_{f}$  is fixed within the inflationary sector, while $N_{*}$  depends on the integrated post-inflationary expansion of the universe, and is thus sensitive to unknown physics down to  LHC scales, and possibly below \cite{Adshead:2010mc,Easther:2011yq}.}
Consequently, the parameter $N_*$ affects the allowed number of fields, causing $N_f$ to have more impact on the  distributions $P(n_s \| N_f)$ and $P(\alpha_s \| N_f)$ than it does for BayesNet A or BayesNet B.
However, at large $N_{f}$ we see behaviour similar to BayesNet A and BayesNet B, since the \emph{a priori} covariance between $N_f$ and $N_*$ vanishes in this limit according to Fig.~\ref{fig:priors_rejsample}.
We conclude that the unknown reheating physics, as encoded in the joint prior $P(N_*, N_f)$, remains vitally important for the predictions of BayesNet C, despite the more complicated dependence on $N_f$ and $\beta$ in Figs.~\ref{fig:DAG_nquad2}~and~\ref{fig:model2_hist}.

\section{Summary \& conclusions}
\label{sect:concl}

In Sect.~\ref{sect:hier} we introduced generative hierarchical models and Bayesian networks.  We emphasized the role played by hyperparameters and hyperpriors in fixing  the predictions of specific inflationary models.  While more layers of hyperparameters increases the variance in the predicted distributions of observables, the results are more robust in that they more fully reflect the unknown physics of inflation. 

In Sect.~\ref{ssect:pheno} we looked at a phenomenologically motivated Bayesian network with minimal dependencies between the fields' initial conditions $\phi_{i,0}$, masses $m_i^2$, the total number of fields, and $N_*$.  We  required that the model gives exactly $N_*$ $e$-folds of expansion and stipulated that the prior probability on the initial conditions is uniformly weighted on the field space regions  satisfying this relationship.  We chose the total number of $e$-folds uniformly from the range $40 \le N_* \le 70$, the number of fields uniformly from the range $1 \le N_f \le 200$, and set the prior probability for the fields' masses to the Mar\v{c}enko-Pastur distribution.  Examining the marginalized distributions $P(n_s)$ and $P(\alpha_s)$ we found that the most important parameter is $N_*$,  indicating that the inherent uncertainty in the predictions of $N_f$--quadratic inflation are far more sensitively dependent on the unknown physics of reheating than on the fields' masses, initial conditions, or the total number of active fields. The dominant source of uncertainty in this model is therefore the same as that which plagues even the simplest single field models.

Section~\ref{ssect:maxent} used the same Bayesian network as Sect.~\ref{ssect:pheno}, but updated the prior probability for the number of fields to the geometric distribution, which maximizes the Shannon entropy, with the \emph{a priori} expected number of fields $\vev{N_f}$ entering as new hyperparameter.  Although this prior substantially reweights the number of active fields in a given inflationary realization,  the model's predictions are indistinguishable from BayesNet A of Sect.~\ref{ssect:pheno} for $\vev{N_f} \gtrsim 20$, but are noticeably different for $\vev{N_f} \ll 20$.

Section~\ref{ssect:hep} explored the dependency on the initial energy scale $\rho_0$ in a model where the fields are required to have sub-Planckian  displacements $|\phi_{i,0}| \ll \mpl$, while varying the expected ratio of axions to the total number moduli fields in the string theory motivated version of $N_f$--quadratic inflation of Ref.~\cite{Easther:2005zr}.  The reduced range for $\phi_{i,0}$ requires more fields to get at least $N_*$ $e$-folds of inflation~\cite{Liddle:1998jc,Dimopoulos:2005ac}.  Since  lower values of $N_*$ need fewer fields to achieve enough inflation, there is a significant coupling between the prior probabilities for $N_*$ and $N_f$,  resulting in  predictions for $n_s$ and $\alpha_s$ differing from BayesNet A or BayesNet B.  Examining the marginalized distributions for $n_s$ and $\alpha_s$, we  still qualitatively identify $N_*$ as the most important parameter for determining the model's prediction, although there are now more complicated interdependencies between $N_*$, $\beta$, and $N_f$.  With a non-trivial prior for $N_f$ and allowing the ratio of axions to the total moduli to vary in the mass distribution, this model has a greater dependence on the total number of fields and masses than BayesNet A or BayesNet B, especially for $N_f \lesssim 600$.  Despite this, we can identify a limiting behavior for large $N_f$ where $N_*$ is again clearly the most important parameter.

While our results are restricted to the case of $N_f$--quadratic inflation, the hierarchical modelling methods that we have used can be applied to any inflation model.  These techniques would be especially well suited to studying models that have non-standard features, such as couplings to other sectors and non-canonical kinetic terms.
Most importantly, the methods developed here will be indispensable for obtaining posterior probabilities on the parameters of complex inflation models with CMB or large scale structure data.

\acknowledgments

We thank Grigor Aslanyan, Michael Betancourt, Mafalda Dias, Jo\~ao Morais, David Seery, Kepa Sousa, and Jonathan White for helpful discussions.
HVP was supported by the European Research Council under the European Community's Seventh Framework Programme (FP7/2007- 2013) / ERC grant agreement no 306478-CosmicDawn. JF was supported by IKERBASQUE, the Basque Foundation for Science and is supported by the ERC Consolidator Grant STRINGFLATION under the HORIZON 2020 contract no. 647995.
The authors acknowledge the use of the New Zealand eScience Infrastructure (NeSI) high-performance computing facilities. New Zealand's national facilities are provided by NeSI and funded jointly by NeSI's collaborator institutions and through the Ministry of Business, Innovation \& Employment's Research Infrastructure programme [{\url{http://www.nesi.org.nz}}].
This work has been facilitated by the Royal Society under their International Exchanges Scheme.
This work was supported in part by National Science Foundation Grant No. PHYS-1066293 and the hospitality of the Aspen Center for Physics. We thank the Galileo Galilei Institute for Theoretical Physics for their hospitality and the INFN for partial support during the completion of this work.

\bibliographystyle{JHEP}
\bibliography{references}

\end{document}